 \documentclass[final,5p,times,twocolumn,authoryear]{elsarticle}

\usepackage{amssymb}
\usepackage{amsmath}
\usepackage{lipsum}
\usepackage{float}
\usepackage{xcolor}

\usepackage{subcaption}

\newcommand{\be}{\begin{eqnarray}}
\newcommand{\ee}{\end{eqnarray}}
\newcommand{\beq}{\begin{equation}}
\newcommand{\eeq}{\end{equation}}
\newcommand{\bemul}{\begin{multline}}
\newcommand{\eemul}{\end{multline}}

\journal{Astroparticle Physics}

\begin{document}

\begin{frontmatter}



\title{Neutrinos from hidden ultraluminous X-ray sources in the Galaxy}


\author[second1,second2]{Lucas Pasquevich}
\author[second1,second2]{Gustavo E. Romero}
\affiliation[second1]{organization={
Instituto Argentino de Radioastronomía (CCT La Plata, CONICET; CICPBA; UNLP)},
            addressline={C.C.5}, 
            city={Villa Elisa},
            postcode={1894}, 
            state={Provincia de Buenos Aires},
            country={Argentina}}   
\affiliation[second2]{organization={
Facultad de Ciencias Astronómicas y Geofísicas, Universidad Nacional de La Plata,  La Plata, Argentina},
            addressline={Paseo del Bosque S/N}, 
            city={La Plata},
            postcode={1900}, 
            state={Provincia de Buenos Aires},
            country={Argentina}}    
\author[first,first2]{Mat\'ias M. Reynoso}
\affiliation[first]{organization={Instituto de Investigaciones Físicas de Mar del Plata (IFIMAR – CONICET), and Departamento de Física, Facultad de Ciencias Exactas y Naturales, Universidad Nacional de Mar del Plata},
            addressline={Funes 3350}, 
            city={Mar del Plata},
            postcode={7600}, 
            state={Provincia de Buenos Aires},
            country={Argentina}}

\affiliation[first2]{organization={Departamento de Física, Facultad de Ciencias Exactas y Naturales, Universidad Nacional de Mar del Plata},
            addressline={Funes 3350}, 
            city={Mar del Plata},
            postcode={7600}, 
            state={Provincia de Buenos Aires},
            country={Argentina}}
            
\begin{abstract}
Ultraluminous X-ray sources (ULXs) are point-like sources that exhibit apparent X-ray luminosities exceeding the Eddington limit for stellar-mass compact objects. A widely accepted interpretation is that these systems are X-ray binaries accreting matter possibly at super-Eddington rates. In this regime, photon trapping inflates the accretion disk, making it geometrically and optically thick. Radiation-driven winds launched from the supercritical disk form funnel-shaped walls along the symmetry axis. While the apparent X-ray luminosity can exceed the Eddington limit due to geometrical beaming within this funnel, a misalignment with the observer's line of sight strongly suppresses the X-ray emission, rendering the ULX electromagnetically obscured.

This work explores the potential for high-energy neutrino production in black hole-hosting ULXs. We model proton acceleration via magnetic reconnection in the region above the super-accreting black hole. Although electromagnetic emission is efficiently absorbed by the dense wind and radiation fields, neutrinos generated from photomeson interactions can escape. Our model self-consistently accounts for energy losses of pions and muons in this environment. The results indicate that misaligned, electromagnetically obscured Galactic ULXs could produce a neutrino flux detectable by instruments like KM3NeT and IceCube within several years of observation.
\end{abstract}




\end{frontmatter}




\section{Introduction} \label{sec:intro}

Galactic sources are known to accelerate cosmic rays up to energies of $\sim$ 1 PeV, the so-called knee of the spectrum. The maximum energy attainable in a given source is governed by its intrinsic properties, such as size, magnetic field strength, age, and acceleration rate, while the large-scale Galactic magnetic field dictates the confinement and propagation of escaping particles \citep{1984ARA&A..22..425H}. When accelerated hadrons interact with ambient matter or radiation fields via inelastic $pp$ or $p\gamma$ collisions, they produce charged pions. These pions decay, yielding high-energy neutrinos and gamma rays with comparable fluxes. In contrast, purely leptonic processes—such as inverse Compton (IC) scattering, synchrotron radiation, and Bremsstrahlung—generate only gamma rays. Consequently, the detection of neutrinos provides unambiguous evidence for hadronic acceleration and offers a unique means of distinguishing between leptonic and hadronic emission mechanisms \citep{GONZALEZGARCIA2009437}. Identifying the hadronic content of Galactic cosmic-ray accelerators therefore requires multimessenger observations combining both gamma-ray and neutrino data.

A growing body of evidence has identified candidate Galactic PeVatrons. For instance, deep imaging of the Galactic Centre region by H.E.S.S. has revealed gamma-ray emission indicative of particle acceleration to PeV energies \citep{Abramowski2016}. Furthermore, wide-field air-shower arrays such as LHAASO, Tibet AS$\gamma$, and HAWC have detected very-high-energy gamma rays extending to hundreds of TeV and up to ~PeV from multiple sources throughout the Galaxy \citep{LHAASO2021, Amenomori2021, Abeysekara2020}. These detections imply the presence of parent particles accelerated to energies $\gtrsim 10$PeV. However, the gamma-ray data alone cannot unambiguously confirm a hadronic origin, as leptonic processes can also produce photons in this energy range. This ambiguity underscores the critical role of neutrinos as a decisive diagnostic tool. Their minimal interaction cross-section allows them to traverse dense astrophysical environments without significant attenuation, making them ideal messengers for probing the sites of cosmic accelerators and directly tracing the relativistic hadronic content at their sources.

The IceCube Collaboration has recently reported evidence for a diffuse neutrino flux originating from the Galactic plane, identified using a template based on \textit{Fermi-LAT} $\pi^0$-decay maps \citep{Gaggero_2015, IceCube_galactic_neutrino_2023, 2024PhRvD.109d3007A}. Despite this significant advance, the specific sources responsible for this Galactic neutrino emission remain largely unidentified.

In a parallel development, the LHAASO Collaboration has detected gamma rays with energies exceeding 100 TeV from several Galactic microquasars, including SS 433, Cygnus X-1, V4641 Sgr, GRS 1915+105, and MAXI J1820+070 \citep{LHAASOCollaboration_BHJet_2024}. The presence of such ultra-high-energy photons implies that these systems are efficient particle accelerators—potentially via shocks or magnetic reconnection—and are viable candidate PeVatrons \citep{Abaroaetal2024, Peretti2025, Wang2025ApJ, Zhang2025arXiv250620193Z}. While the hadronic origin of this radiation is an active area of investigation \citep{2025arXiv250622550C}, these findings highlight the potential of microquasars as neutrino sources and create new opportunities for multimessenger studies of Galactic high-energy phenomena \citep{Romero2001A&A, Anchordoqui2003ApJ, 2003A&A...410L...1R, 2006PhRvD..73f3012C}. This recent focus builds upon a foundation of earlier theoretical work that explored the mechanisms and detectability of high-energy neutrino emission from accreting compact objects, including neutron stars and microquasars \citep{Distefano_2002, Torres-Romero-Mirabel2005ChJAS, Romero-Orellana2005A&A, Orellana_2007A&A, 2008MNRAS.387.1745R, Vila_Romero_2010, reynosocarulli2019, sudoh2020, kimura2020}.

The currently most widely accepted interpretation of ultraluminous X-ray sources (ULXs) is that they are X-ray binaries hosting a stellar-mass compact object accreting at super-Eddington rates \citep[see][for reviews]{Fabrika_rev_2021AstBu,King_rev_2023NewAR..9601672K}. These systems exhibit X-ray luminosities exceeding $10^{39}$ erg s$^{-1}$, above the Eddington luminosity of a stellar black hole (BH) with $M_{\rm BH}\lesssim 10M_\odot$, under the assumption of isotropic emission. The Eddington luminosity is defined as 
\begin{equation}
    L_{\rm Edd}= 1.26 \times 10^{38} \left( \frac{M_{\rm BH}}{\rm M_\odot} \right)\,\,\rm{erg\,s^{-1}},
\end{equation}
and the corresponding Eddington accretion rate as
\begin{equation}
    \dot{M}_{\rm Edd} = \frac{L_{\rm Edd}}{\eta \, c^2},
\end{equation}
where $c$ is the speed of light and $\eta\sim0.1$ is the accretion efficiency. 

Sources with accretion rates exceeding the Eddington limit ($\dot{M}> \dot{M}_{\rm Edd}$) are in a supercritical regime. In this context, the standard disk model of \cite{shakura1973} breaks down and the disk becomes optically and geometrically thick inside a critical radius $r_{\rm crit}$. The luminosity saturates at the $L_{\rm Edd}$, while the excess material is expelled in the form of powerful radiation-driven winds that regulate the accretion rate onto the BH at $\dot{M}_{\rm Edd}$ \citep[e.g.,][]{lipunova1999,fukue2004,ohsuga2005,akizuki2006,Abaroaetal2024,Abaroa&Romero2024smbh}.

The optically thick wind forms the walls of a funnel with a semi-opening angle $\vartheta$, which exposes the innermost region of the supercritical disk. When an observer's line of sight lies within this funnel, the inner disk is visible as a powerful X-ray source whose apparent luminosity is amplified by geometric beaming. Conversely, if the system is misaligned with the observer, the X-ray emission is strongly suppressed, potentially rendering the source undetectable in this energy band.

Supercritical accretion occurs in a variety of astrophysical systems, including narrow-line Seyfert 1 galaxies, luminous quasars, and transient episodes associated with tidal disruption events. It is also present in X-ray binaries such as SS-433 and Cyg X-3, and in the transient Galactic ULX pulsar (PULX) Swift J0243.6+6124 \citep{Swift_2019Apj,Swift_2022ApJ,Swift_2024A&A}.

The non-thermal component observed in the spectra of some black-hole ULXs potentially originates from both photohadronic interactions and leptonic Comptonization of the disk's X-ray radiation \citep[e.g.,][]{Cruz-Sanchez2025A&A}. In such environments, relativistic protons produce electron-positron pairs via the Bethe-Heitler mechanism \citep{bethe_heitler_1934}; these pairs subsequently upscatter photons through interactions with ambient radiation, emitting hard X-rays and soft gamma-rays \citep[see][]{Romero-Pasqa-Abaroa2025}. Furthermore, the acceleration of hadrons to PeV energies in these systems leads to charged pion production, whose decay could yield a non-negligible flux of high-energy neutrinos, establishing ULXs as potential contributors to the Galactic neutrino background.

Neutrino production has also been predicted for the neutron-star subclass of ULXs, particularly in the MeV energy range \citep{Asthana_2023, Mushtukov2025}. Recent models suggest that their neutrino emission may have been substantially underestimated. Expanding to higher energies, \cite{NSULX_neutrinos_2025} recently proposed that inelastic collisions between protons accelerated to TeV energies in the magnetosphere of a magnetized neutron star ($B \sim 10^{12}$ G) and protons from the accretion disk generate significant neutrino fluxes extending up to several tens of TeV. This mechanism is consistent with earlier work on hadronic processes in similar environments \citep{Romero2001A&A, Anchordoqui2003ApJ}.

This paper investigates neutrino production through hadronic processes in Galactic ULXs harboring stellar-mass black holes. We specifically focus on the scenario where these sources are misaligned, rendering them obscured in the X-ray band while allowing a portion of their higher-energy gamma-ray and neutrino emission to escape. If these systems prove to be efficient hadronic accelerators, our models indicate they would produce a neutrino flux detectable by current and next-generation observatories such as IceCube, IceCube-Gen2, and KM3NeT.

The structure of the paper is as follows. Section 2 details the development of a physical model for two representative ULXs with different accretion rates. Within this framework, we characterize the magnetic confinement region via the magnetization parameter ($\sigma$) and explore a range of spectral indices for hadron acceleration and injection. We present the resulting hadron distributions and, in Section \ref{sec: particle distribution}, compute the production of secondary particles (pions and muons), tracking their energy evolution through interactions with ambient fields. The corresponding neutrino fluxes are presented in Section \ref{sec: neutrino fluxes}. We discuss the model's assumptions and the prospects for detecting neutrinos from Galactic ULXs in Section \ref{sec: discussion}. Finally, we summarize our work and present the principal conclusions in Section \ref{sec: conclusion}.
                             
\section{Basics of the model} \label{sec:model}

\begin{figure}[h!]
    \centering   \includegraphics[width=0.9\linewidth]{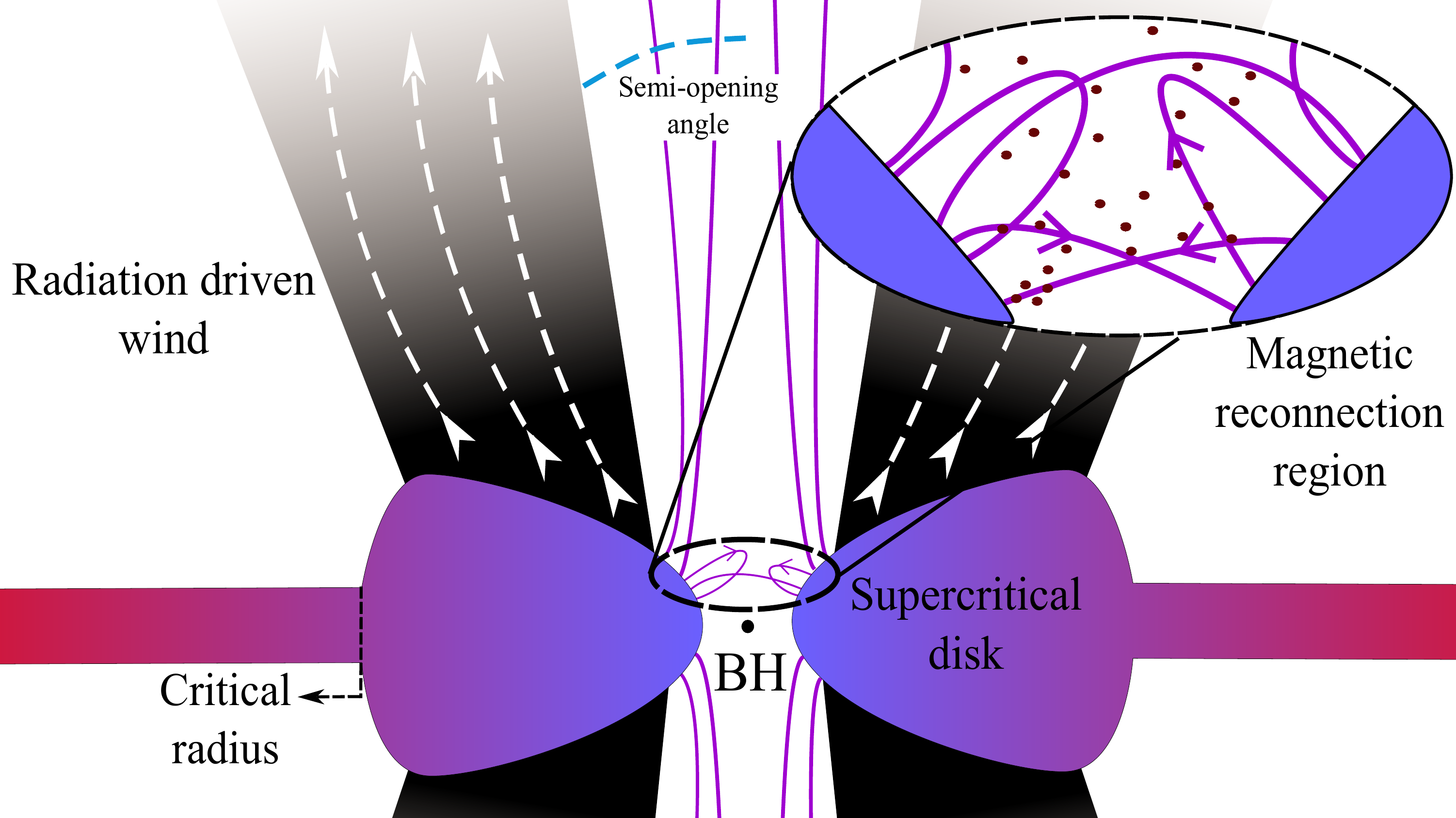}
 \caption{Schematic diagram of a ULX system. Powerful winds from the supercritical disk form a funnel around the black hole. Inside this funnel, a magnetically dominated region enables particle acceleration via magnetic reconnection.}\label{fig.sketch}
\end{figure}

We assume the existence of a particle acceleration region located inside the funnel and above the BH, where hadrons can be accelerated to high energies. A schematic representation of the system is shown in Fig. \ref{fig.sketch}. The acceleration region (dashed area) is located a short distance, $z_{\rm acc}$, from the BH along the symmetry axis of the system. Within this region, particles are magnetically confined, and both protons and electrons are accelerated via magnetic reconnection. The reconnecting field is associated with the accreting flow, where, at small scales, turbulence induces fast reconnection between small accreting magnetic loops and the non-dipolar field, analogous to processes observed in the solar corona \citep[e.g.][]{El2022A&A,Kimura2022ApJ,Karavola2025JCAP.}. Magnetic confinement is expected on spatial scales comparable to the BH size \citep{magReconnectionAGN_Liu2003,CoronaCyg1_Romero2014,AGNNuSTAR_Fabian2015,AstroparticlesCoronae_Fang2024} where non-dipolar field is accreted and has a high rate of magnetic reconnection; at larger distances above the disk, the dominant magnetic field is expected to become more ordered and predominantly poloidal, since it is originated from currents in the disk \citep{Rothstein_Advection_MagField_2008,Beloborodov_MagRec_2017,Ide_Dynamo_2024}.

We consider two accretion scenarios: models labeled $A$ correspond to a ULX with an accretion rate of $\dot{m}=10$, whereas models $B$ represent a hyperaccreting ULX with $\dot{m}=10^3$ \citep{King2008}, where $\dot{m}=\dot{M}/\dot{M}_{\rm Edd}$ is the accretion rate in Eddington units. These values are intended to bracket the relevant range for super-Eddington accretion, from a moderately super-Eddington regime ($A$) to a hypercritical one ($B$), intermediate accretion rates are expected to yield results that smoothly interpolate between these two cases. In both scenarios, the disk is thick inside the critical radius
\begin{equation}
r_{\rm crit} = 1.95 \, \dot{m} \, r_{\rm g},
\end{equation}
beyond which it becomes geometrically thin and optically thick. Hence, $r_{\mathrm{crit}, \,A} = 19.5\,  r_{\rm g}$ and $r_{\mathrm{crit},\, B} = 1.95 \times 10^3  r_{\rm g}$, where $r_{\rm g} = G M_{\rm BH} / c^2$ is the gravitational radius and $G$ is the gravitational constant.

We adopt the magnetized, super-Eddington accretion disk solution of \citet{fukue2004} and \citet{akizuki2006}. The physical properties of the disk are characterized by a set of global parameters. We adopt the representative values $\alpha=0.01$ for the viscosity parameter, $\beta \equiv p_{\rm mag}/p_{\rm gas}=5$ for the inner-disk magnetization, $\gamma=4/3$ for the adiabatic index, and $f=0.5$ for the advection parameter. The toroidal magnetic field is defined as \citep{akizuki2006}:
\begin{equation}
    B_{\phi}(r)=\sqrt{4\pi \Sigma_0 G M_{\rm BH} \frac
{\beta c_3}{\sqrt{(1+\beta)c_3}}}\,r^{s/2-1},
\end{equation}
where $\Sigma_0$ is the surface density of the disk, $s=1/2$ parametrizes the mass loss through the winds, and $c_3(\alpha,\beta,\gamma,f,s)$ is a dimensionless coefficient of the model that depends on the adopted disk parameters (see \citealt{fukue2004} for the full expression). This model yields an inner-disk temperature of $T \approx 10^{7}$ K for both configurations $A$ and $B$. The magnetic field, however, differs significantly: $B \simeq 5 \times 10^{7}$ G for configuration $A$ and $B \simeq 5 \times 10^{6}$ G for $B$.

As discussed in the Introduction, the disk-driven wind expelled from the disk forms a funnel with a semi-opening angle $\vartheta$. The value of $\vartheta$ depends on many factors, from the disk's angular momentum to the black hole mass. From an observational point of view, it is clear that this angle cannot be too large, as otherwise the geometrical collimation of the X-ray flux would not occur. We adopt $\vartheta_A=15^\circ$ and $\vartheta_B=10^\circ$, taking the value $\vartheta\approx15^\circ$ inferred for the supercritical source Cyg X-3 as a reference (e.g., \citealt{Veledina_etal_cygx3_2024NatAs}). 

ULX's X-ray spectra are often described by a thermal disk component plus a harder power-law tail \citep[see][and references therein]{Fabrika_rev_2021AstBu}. The detection of thermal emission from the inner disk indicates that the system is viewed at a relatively low inclination. This means that the line of sight passes through the funnel and intercepts the innermost regions of the disk. Therefore, the funnel must be optically thin, and the optical depth is constrained by \citep{ObservationalBHWinds_Fukue2009,photosphereThermalization_Tomida2015}:
\begin{equation}
\tau = \int^\infty_{0} \gamma_{\rm gas}(1-\beta_{\rm gas} \cos{\vartheta}) \, \kappa_{\rm co} \,\rho_{\rm co} \, {\rm d}z < 1,
\end{equation}
where $\rho_{\rm co}\approx \rho_{\rm gas}$ and $\kappa_{\rm co}\approx\kappa=\sigma_{\rm T}/m_{\rm p}$ are the density and opacity in the comoving frame, respectively, $\sigma_{\rm T}$ is the Thomson cross section, and $\beta_{\rm gas} = v_{\rm gas}/c$ is the bulk velocity of the gas in units of the speed of light and $\gamma_{\rm gas}$ is the corresponding Lorentz factor. The above condition establishes an upper limit on the gas density, $\rho_{\rm gas}$, for the funnel to remain optically thin. This is the maximum limit; actual sources are expected to have densities below this value.


\begin{table*}[h!]
{\small 
\caption{Parameter configurations for the ULX models}\label{table:params}
\centering                                      
\begin{tabular}{l c c c c c c c}          
\hline                     
parameter [unit] &  description &    $A_1$ &  $A_2$ & $A_3$ &  $B_1$ &  $B_2$ &  $B_3$ \\    
\hline                                   
    $M_{\rm BH} {[M_\odot]}$ & black hole mass  & $10$& $10$& $10$& $10$ & $10$ & $10$ \\
    $\dot{m}\, {[\dot{M}_{\rm Edd}]}$ & accretion rate & $10$ &  $10$ & $10$ & $10^3$ & $10^3$ & $10^3$ \\
   $T_{\rm d} $ [K] & inner disk temperature & $ 10^7$& $10^7$& $10^7$& $10^7$ & $10^7$& $10^7$\\
   $z_{\rm acc}\ [r_g]$  &acceleration position & $10$& $10$& $10$& $10$  & $10$& $10$\\
    $\vartheta \, [^\circ]$ & funnel semi-opening angle & $15$ &  $15$ &  $15$ &  $10$&  $10$ &  $10$\\
   $T_{\rm w} $ [K] & wind temperature & $4\times 10^5$& $4\times 10^5$& $4\times 10^5$& $1.2\times 10^4$& $1.2\times 10^4$& $1.2\times 10^4$ \\      
    $B$ \ [G] &  magnetic field & $5\times 10^7$ & $5\times 10^7$ & $5\times 10^7$ & $5\times 10^6$ & $5\times 10^6$ & $5\times 10^6$\\
    $n_{\rm p} \ [{\rm cm^{-3}}]$  &  gas density  & $1.3\times 10^{16}$ & $2.6\times10^{15}$& $4.4\times 10^{14}$ & $1.3\times10^{14}$& $2.6\times 10^{13}$ & $4.4\times 10^{12}$  \\
$\Gamma$ & injection index & $2$& $1.5$& $1$& $2$  & $1.5$& $1$ \\
\hline
\end{tabular}
}
\end{table*}

\begin{figure*}[h!]                            
\centering
  \centering
    \begin{subfigure}[t]{0.49\textwidth}
        \centering            \includegraphics[width=0.5\linewidth,trim= 130 30 120 20]{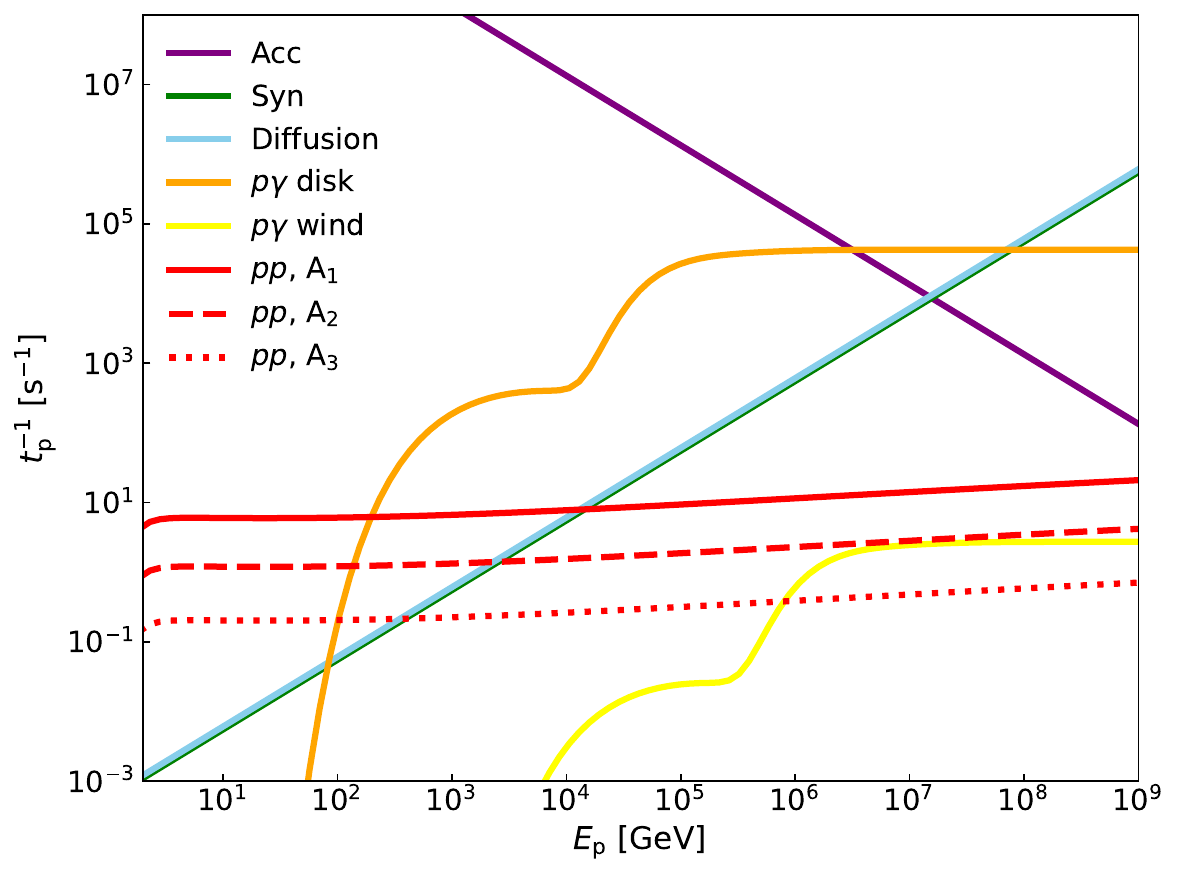} 
    \end{subfigure}
    \hfill
    \begin{subfigure}[t]{0.49\textwidth}
        \centering
    \includegraphics[width=0.5\linewidth,trim= 130 30 120 30]{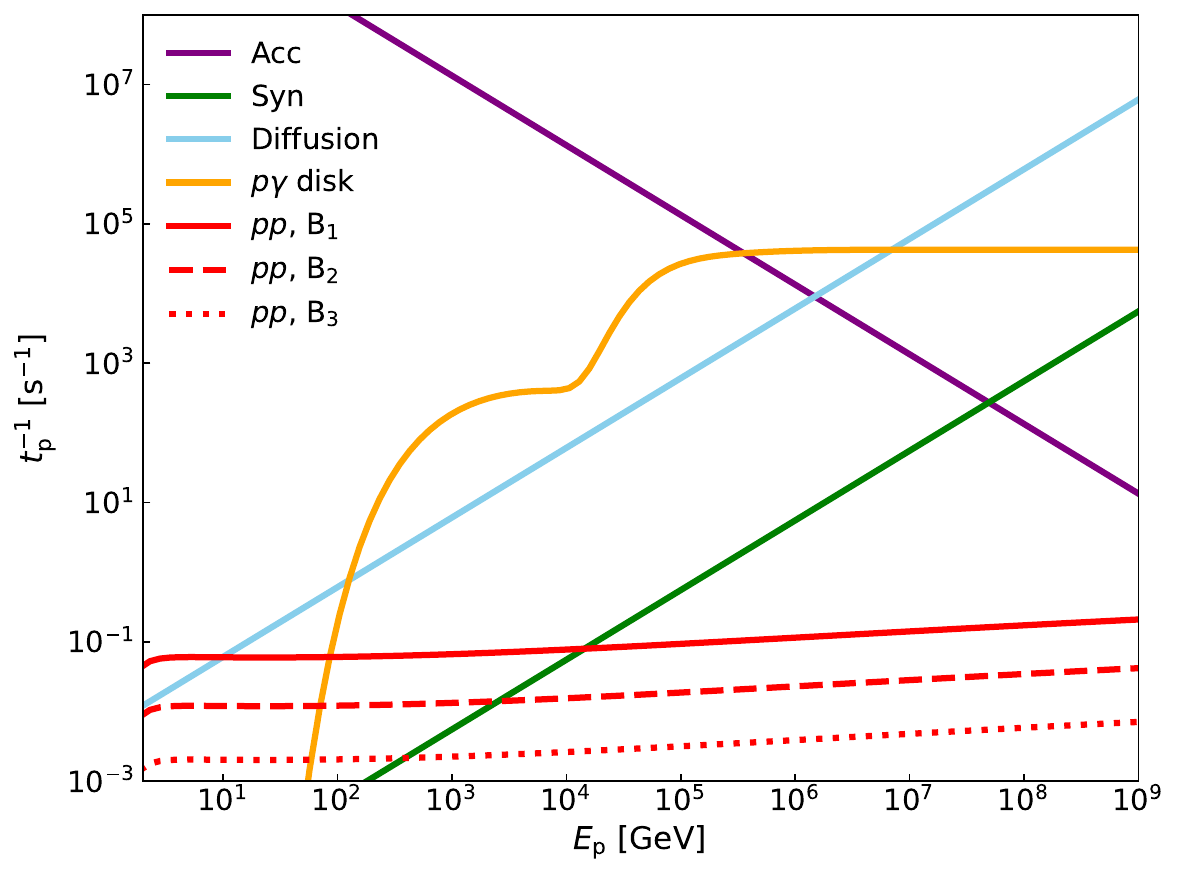} 
    \end{subfigure}
   
\caption{Proton acceleration, cooling, and escape rates for the different parameter sets. Left: Scenario with an accretion rate of $\dot{m}=10$ for different densities ($A_1$, $A_2$, and $A_3$). At low energies, $pp$ interactions dominate the cooling, while above $\sim 100$~GeV, $p\gamma$ interactions become the main channel. The maximum proton energy reaches the PeV range. Right: Scenario with $\dot{m} = 1000$ for a range of densities ($B_1$, $B_2$, and $B_3$). As in the left panel, $pp$ interactions dominate at low energies and $p\gamma$ processes take over above $\sim 100$GeV. However, due to the strong cooling, the maximum energy is limited to $\sim 100$TeV.}

 \label{fig:p-rates}
\end{figure*}
%


We assume uniform physical conditions inside the acceleration region, which has a size of $\Delta z = 0.1 z_{\rm acc}$. Protons are confined by the strong, turbulent magnetic field near the black hole and interact primarily with the X-ray photon field from the inner disk.

The acceleration rate for a charged particle in a magnetic field $B$ is given by \citep[see][]{drury1983,aharonian2004}:
\begin{equation}
t_{\mathrm{acc}}^{-1} = \frac{\eta e c B}{E},
\label{equation_AccelerationRate}
\end{equation}
where $\eta$ is the acceleration efficiency. In the case of magnetic reconnection, this efficiency is \citep{2011ApJ...738..115D,2012A&A...542A...7V} 
\begin{equation}
\eta_{\rm rec} \sim 0.1 \frac{r_{\rm g} c}{D(E)} \left( \frac{v_{\rm rec}}{c} \right)^2,
\end{equation}
where $D(E) = r_{\rm gy} c / 3$ is the Bohm diffusion coefficient and $r_{\rm gy} = E / (eB)$ is the gyroradius.
Under strong magnetic fields, as is the case here, the reconnection velocity is similar to the Alfvén velocity, $v_{\rm rec} \sim v_{\rm A}$. Under these conditions, $v_{\rm A} \sim c$, leading to $\eta_{\rm rec} \approx 0.3$. This regime is valid in a compact, magnetically confined region near the BH, where magnetic loops and multipolar fields originated in the inner disk are present. At greater heights above the disk, the strength of the magnetic field decreases, which reduces acceleration efficiency and maximum particle energies. Other acceleration mechanisms, such as shock acceleration, may then become dominant. 

The action of a diffusive particle-acceleration mechanism naturally leads to a power-law energy spectrum \citep[e.g.][]{drury1983, Protheroe_1999}. In highly magnetized environments, this mechanism is expected to be mediated not by shock waves but by turbulent magnetic reconnection. Numerical studies of turbulent reconnection show that power laws and acceleration efficiencies are comparable to, or even larger than, those obtained in shocks  \citep{Kowal_Magnetohydrodynamic_Simu_2011,Lazarian_FastMagRec_2011, Lazarian_MagneticReconnection_2012, Sironi_RelativisticRec_2014, Kagan_Magnetic_Rec_2015, Beresnyak_MagneticAcceleration_2016, del_Valle_FermiFirstOrder_2016}. 

Motivated by these results, under the assumptions of homogeneous and isotropic injection, we parameterize the proton injection spectrum as a power law with an exponential cutoff:
\begin{equation}
Q_{\rm p}(E_{\rm p})= \frac{\mathrm{d}\mathcal{N}_{\rm p}}{\mathrm{d}E_{\rm p} \, \mathrm{d}\Omega \, \mathrm{d}V \, \mathrm{d}t}=K_{\rm p}E_{\rm p}^{-\Gamma} \exp\left(-\frac{E_{\rm p}}{E_{{\rm p},\,\rm max}}\right),
\end{equation}
where $Q_{\rm p}$ is the number of non-thermal particles injected per unit energy , solid angle, volume, and time. Here, $d\mathcal{N}_p$ denotes the number of protons injected during a time interval $dt$ within a volume element $dV$, with energies between  $E_{\rm p}$ and $E_p+dE_p$ and momenta directed within a solid-angle element $d\Omega$. On the right side, $K_{\rm p}$ is a normalization constant, $\Gamma$ is the spectral index, and $E_{{\rm p}, \,\rm max}$ is the maximum proton energy. This maximum energy is determined by balancing the acceleration rate against the total rate of energy losses and escape. The normalization constant $K_{\rm p}$ is determined by relating it to the power injected into protons, $L_{\rm p}$:
\begin{equation}
    L_{\rm p}=\int \mathrm{d}V \int\mathrm{d}\Omega\int \mathrm{d}E_{\rm p}\, E_{\rm p}\,Q_{\rm p}(E_{\rm p}).
\end{equation}
Using the above expression for $Q_{\rm p}$, we obtain
\begin{equation}
K_{\rm p}=\frac{L_{\rm p}}{4\pi \, \Delta V \int_{E_{\rm min}}^{\infty} E_{\rm p}^{1-\Gamma}  \exp\left(-\frac{E_{\rm p}}{E_{{\rm p}, \rm max}}\right)  \mathrm{d}E_{\rm p} },
\end{equation}
where the interaction volume is $\Delta V=\Delta z_{\rm acc} \, \pi  r_{\rm f}^{2}$. We take the transverse size $r_{\rm f}=\tan(\vartheta)r_{\rm acc}$ to be determined by the funnel radius at the height $z_{\rm acc}$.

The spectral index $\Gamma$ depends on the plasma magnetization \citep{2015SSRv..191..545K}. A plasma is magnetically dominated if
\begin{equation}
\sigma_{\rm mag}\approx\frac{B^2}{4\pi\, n\, m \,c^2 }>1,
\end{equation}
where $n$ is the number density. Numerical simulations suggest $\Gamma\approx 2$ for a moderate magnetization $\sigma\sim 10$, $\Gamma\approx 1.5$ for $\sigma\approx 50$, and $\Gamma\approx 1$ for highly magnetized flows $\sigma\sim 300$ \citep{Guo_2014,Sironi_2014,Werner_2016}. We therefore explore scenarios with $\Gamma= 2,1.5, {\rm and} \,1$, labeling configurations by $i=1,2,3$ for $A_i$ and $B_i$. The values of the main parameters for the model are shown in Table \ref{table:params}, which presents all six configurations. In the one-zone approximation, which assumes homogeneity and isotropy, these parameters are constant and represent local quantities that characterize the compact region where particles are magnetically confined.

In all configurations, we assume that protons carry more power than electrons, such that $L_{\rm p} = a L_{\rm e}$ with $a = 1000$\footnote{\textbf{The way energy is distributed among hadrons and leptons is not well understood. In cosmic rays, $a=100$ is observed. However, in a strongly magnetized regime, the synchrotron losses of electrons are catastrophic, so the scenario is expected to be highly proton-dominated.}}. Since our focus is on neutrino production processes, we do not model the radiation from primary electrons. For a discussion of the latter, we refer the reader to \citet{Romero-Pasqa-Abaroa2025}.

\begin{figure*}[t]                            
\centering
\  \centering
    \begin{subfigure}[t]{0.49\textwidth}
        \centering            \includegraphics[width=0.5\linewidth,trim= 110 30 110 20]{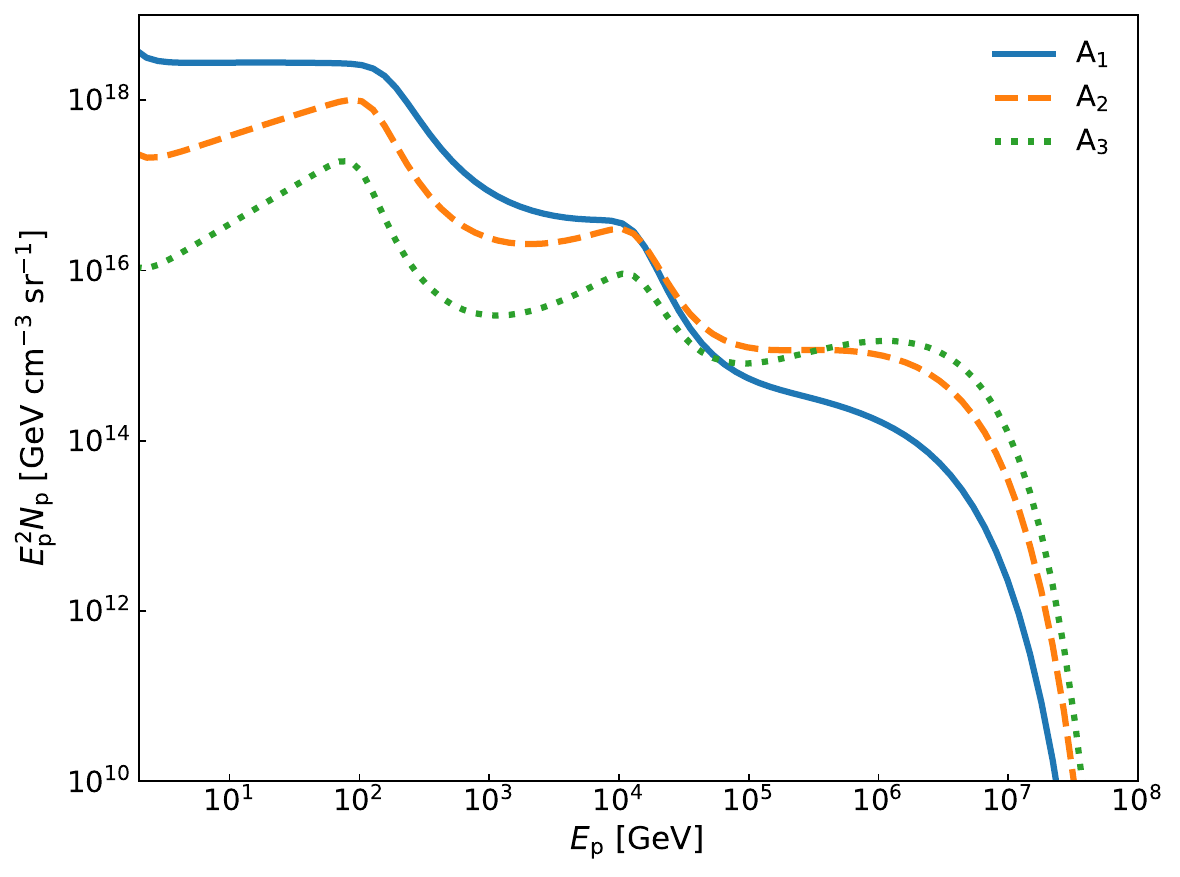} 
    \end{subfigure}
    \hfill
    \begin{subfigure}[t]{0.49\textwidth}
        \centering
    \includegraphics[width=0.5\linewidth,trim= 110 30 110 30]{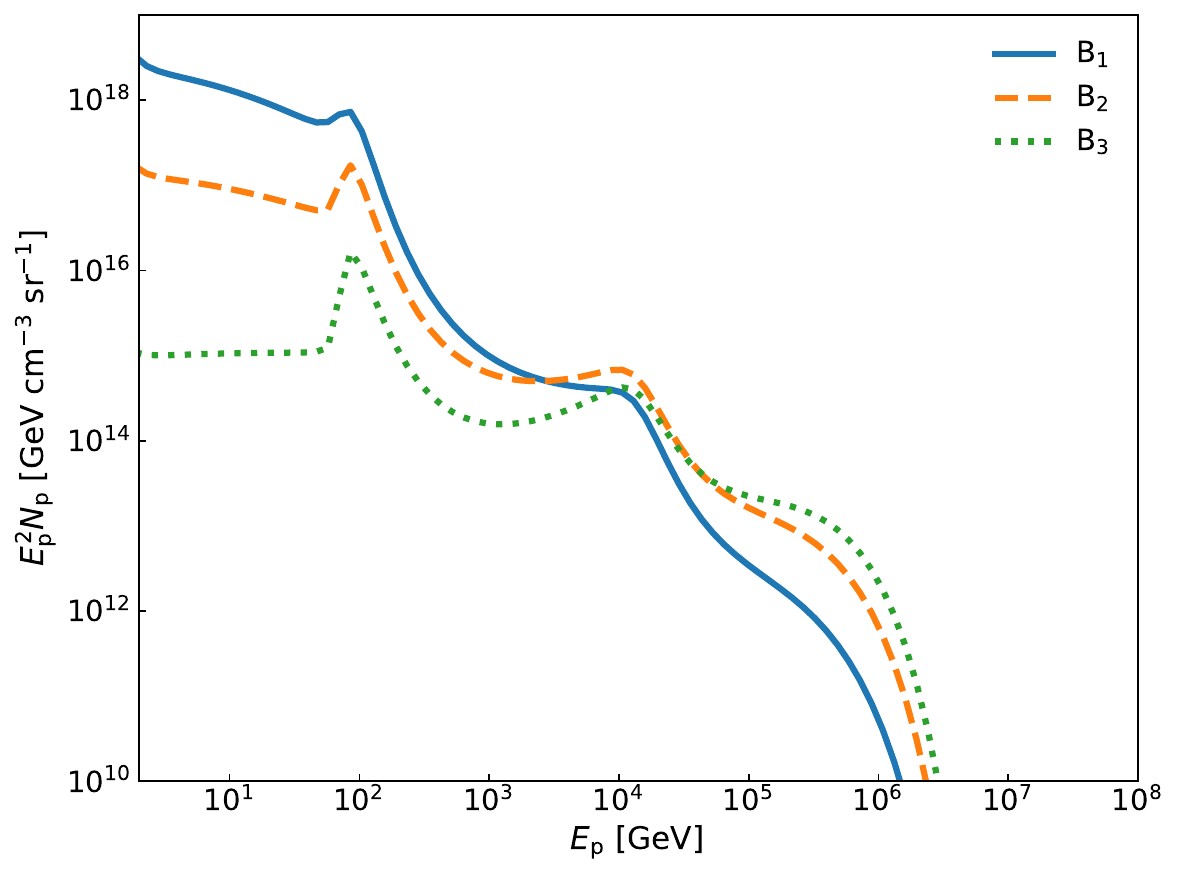} 
    \end{subfigure}
 \caption{Distributions of relativistic protons for scenarios $A_i$ (left) and $B_i$ (right), for different injection spectral indices $\Gamma = 2$, $1.5$, and $1$, corresponding to configurations $i = 1, 2, 3$.
}
\label{fig:p-distributions}
\end{figure*}
We assume that 10\% of the magnetic power available in the reconnection region ($L_{\rm mag}$) is converted into accelerated particles \citep{2011ApJ...738..115D}, such that:
\begin{equation}
L_{\rm p} \simeq 0.1 L_{\rm mag},
\qquad \text{where} \qquad
L_{\rm mag} \approx \frac{B^2 c z_{\rm acc}^2 \tan^2 \vartheta}{2}.
\end{equation}
Here, we have estimated the area of the reconnection zone as $\sim \pi (z_{\rm acc} \tan \vartheta)^2$. Using this formulation, we obtain a proton power of $
L_{\rm p} \simeq 5.9 \times 10^{37} \; {\rm erg \, s^{-1}}$
for configurations $A_i$, and
$L_{\rm p} \simeq 6.2 \times 10^{34} \; {\rm erg \, s^{-1}}$ for configurations $B_i$.

We also account for the potential escape of particles from the acceleration region by calculating the Bohm diffusion timescale $T_{\rm diff} = {3 (\Delta z)^2}/{2 r_{\rm gy} c}$.

Figure \ref{fig:p-rates} shows the proton energy loss rates for $pp$ collisions, $p\gamma$ interactions with thermal photons from the disk and wind, synchrotron cooling, and diffusive escape for configurations $A_i$ and $B_i$. These rates are calculated using standard formulae from \cite{begelman1990,kelner2008}, and \cite{romero2008}. At low energies, $pp$ interactions dominate. At energies above $\sim 100$ GeV, $p\gamma$ interactions become the dominant process in both scenarios. In configurations $B_i$, both the magnetic field strength and the wind temperature are lower than in configurations $A_i$. This implies lower acceleration and synchrotron cooling rates; $p\gamma$ interactions with wind photons also become negligible. The intersection between the acceleration and total energy loss rates occurs at $E_{{\rm p},\, {\rm max}}\sim 1$ PeV for $A_i$ and at $ \sim 100$ TeV for $B_i$.

\begin{figure*}[h!]                            
\centering
\  \centering
    \begin{subfigure}[t]{0.49\textwidth}
        \centering            \includegraphics[width=0.5\linewidth,trim= 125 30 125 20]{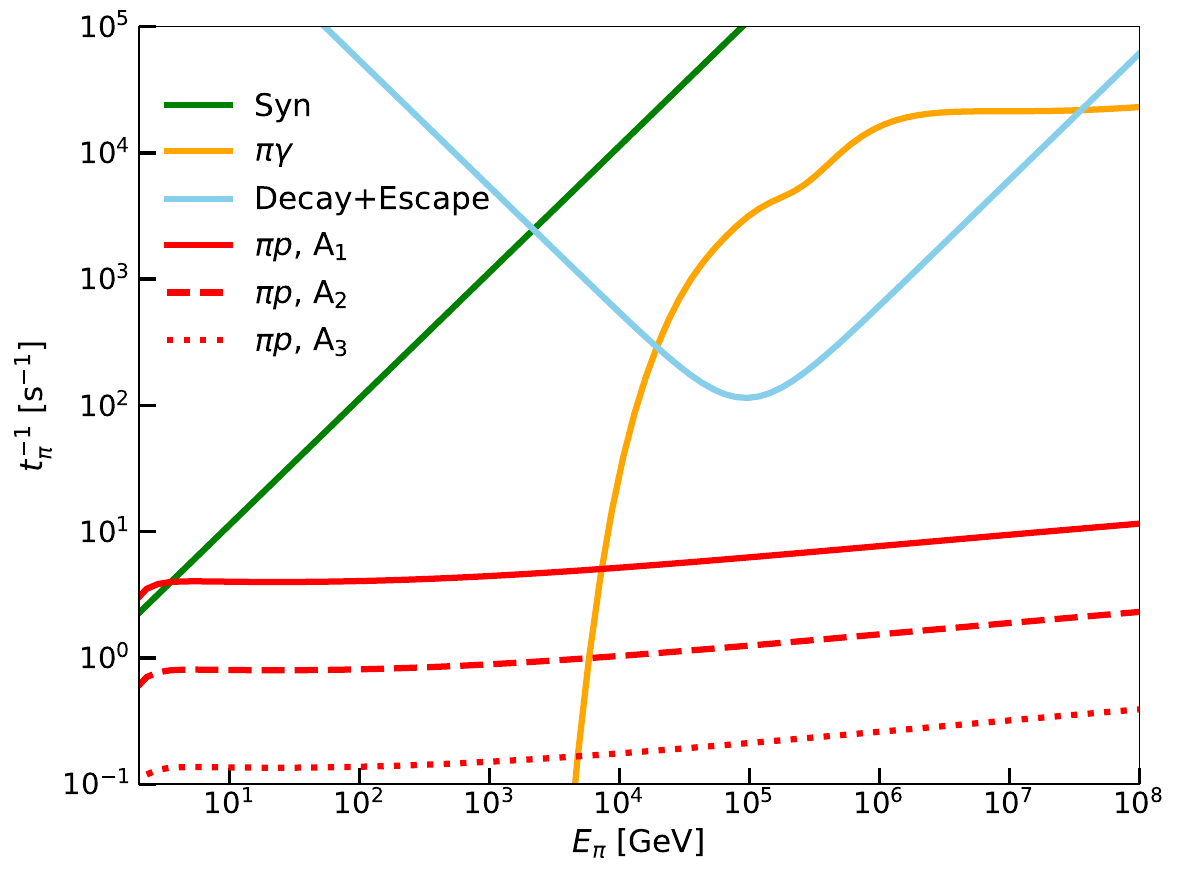} 
    \end{subfigure}
    \hfill
    \begin{subfigure}[t]{0.49\textwidth}
        \centering
    \includegraphics[width=0.5\linewidth,trim= 125 30 125 30]{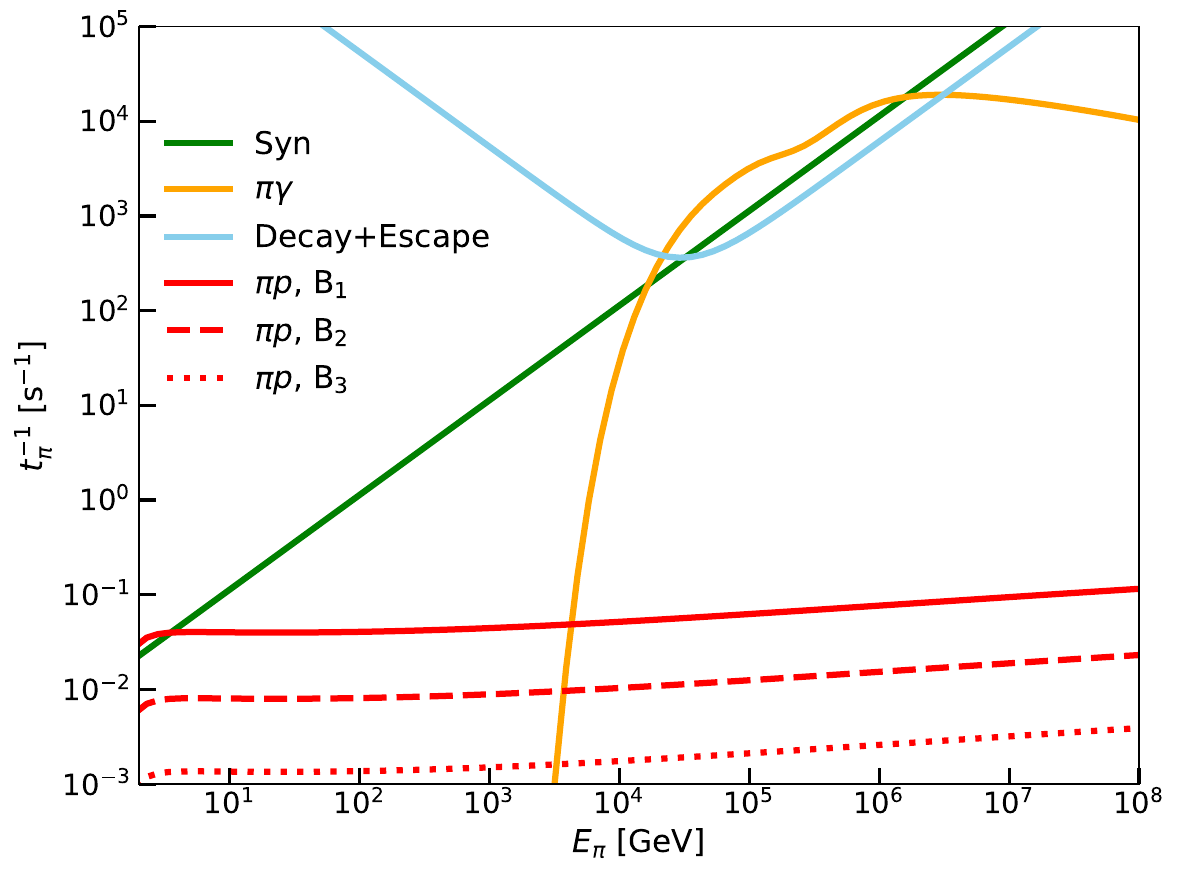} 
    \end{subfigure}
 \caption{Pion cooling, decay, and escape rates for the different parameter sets. We show $\pi\gamma$ interactions with disk photons and $\pi p$ interactions, computed for the three matter densities considered in each set $i$.
Left: $A_i$ scenarios, where particle decay dominates at low energies, while synchrotron losses dominate above $\sim\mathrm{TeV}$.
Right: $B_i$ scenarios, where particle decay dominates at low energies, while $\pi\gamma$ interactions and synchrotron losses dominate above $\sim 10\,\mathrm{TeV}$.}
\label{fig:pi-rates}
\end{figure*}

\begin{figure*}[h!]                            
\centering
\  \centering
    \begin{subfigure}[t]{0.49\textwidth}
        \centering            \includegraphics[width=0.5\linewidth,trim= 115 30 115 20]{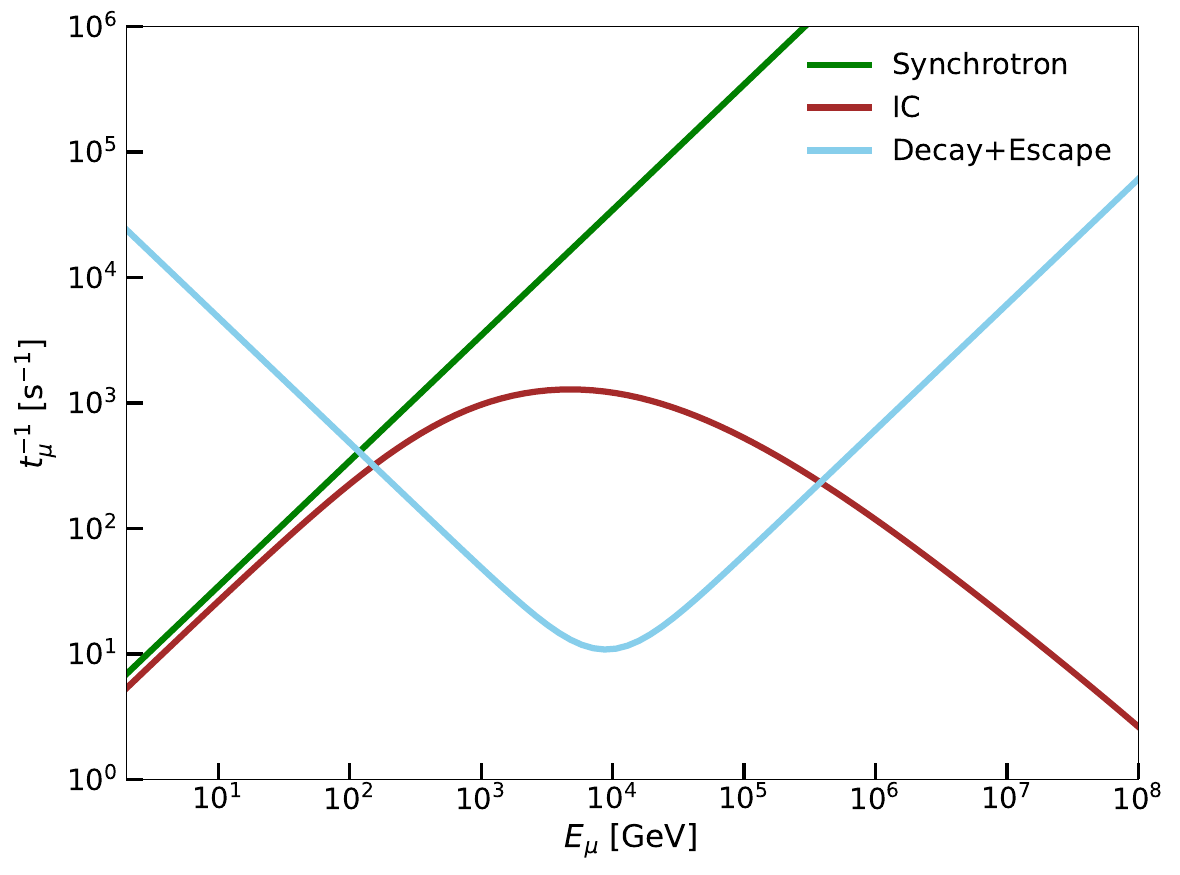} 
    \end{subfigure}
    \hfill
    \begin{subfigure}[t]{0.49\textwidth}
        \centering
    \includegraphics[width=0.5\linewidth,trim= 115 30 115 30]{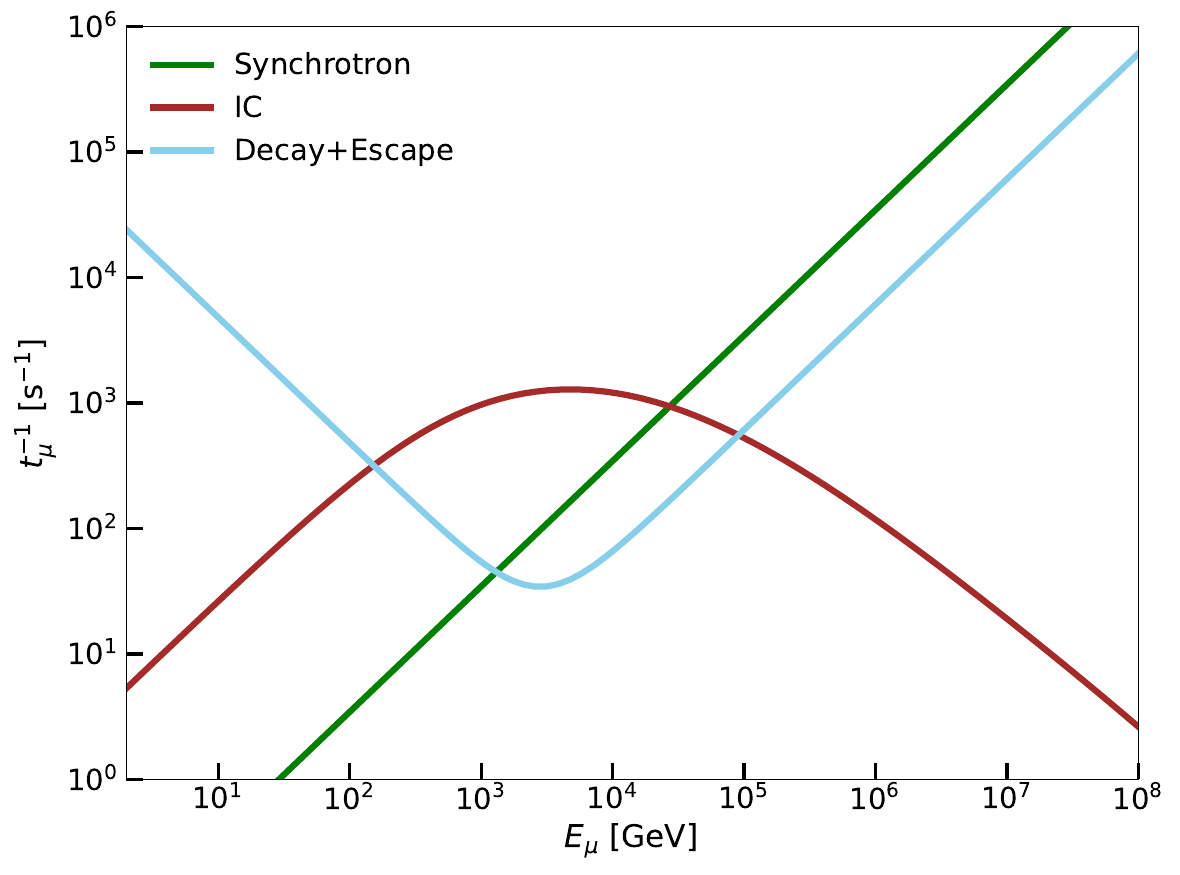} 
    \end{subfigure}
 \caption{Muon cooling, decay, and escape rates for the different scenarios. For the IC interaction, we only consider interactions with the X-photon field of the disk, and we neglect interactions with matter. Left: $A_i$ scenarios, where particle decay dominates at low energies, while synchrotron losses become dominant above $\sim 100$ GeV.
Right: $B_i$ scenarios, where decay dominates at low energies, while IC and synchrotron losses dominate above $\sim 100$ GeV.}
\label{fig:mu-rates}
\end{figure*}

\section{Particle distributions} \label{sec: particle distribution}
\begin{figure*}[t]                            
\centering
\  \centering
    \begin{subfigure}[t]{0.49\textwidth}
        \centering            \includegraphics[width=0.5\linewidth,trim= 120 30 120 20]{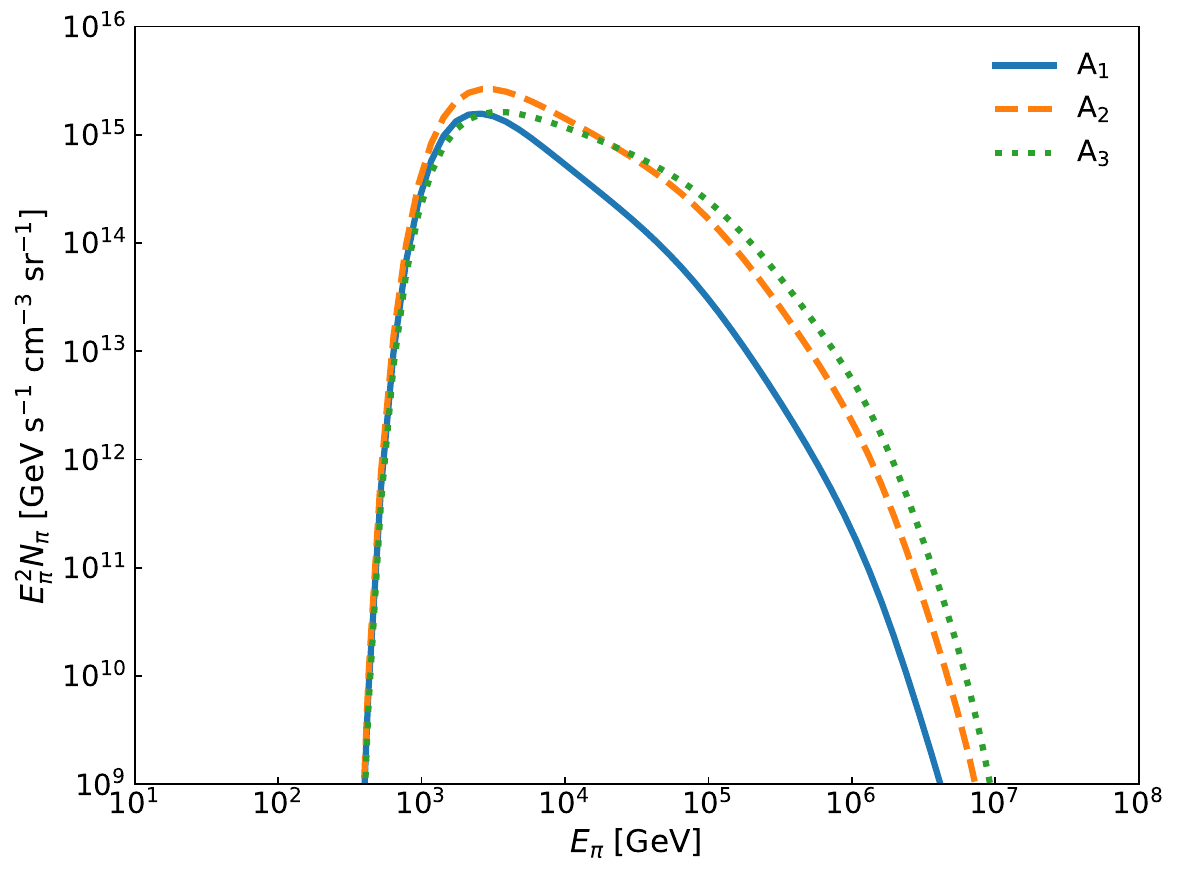} 
    \end{subfigure}
    \hfill
    \begin{subfigure}[t]{0.49\textwidth}
        \centering
    \includegraphics[width=0.5\linewidth,trim= 120 30 120 30]{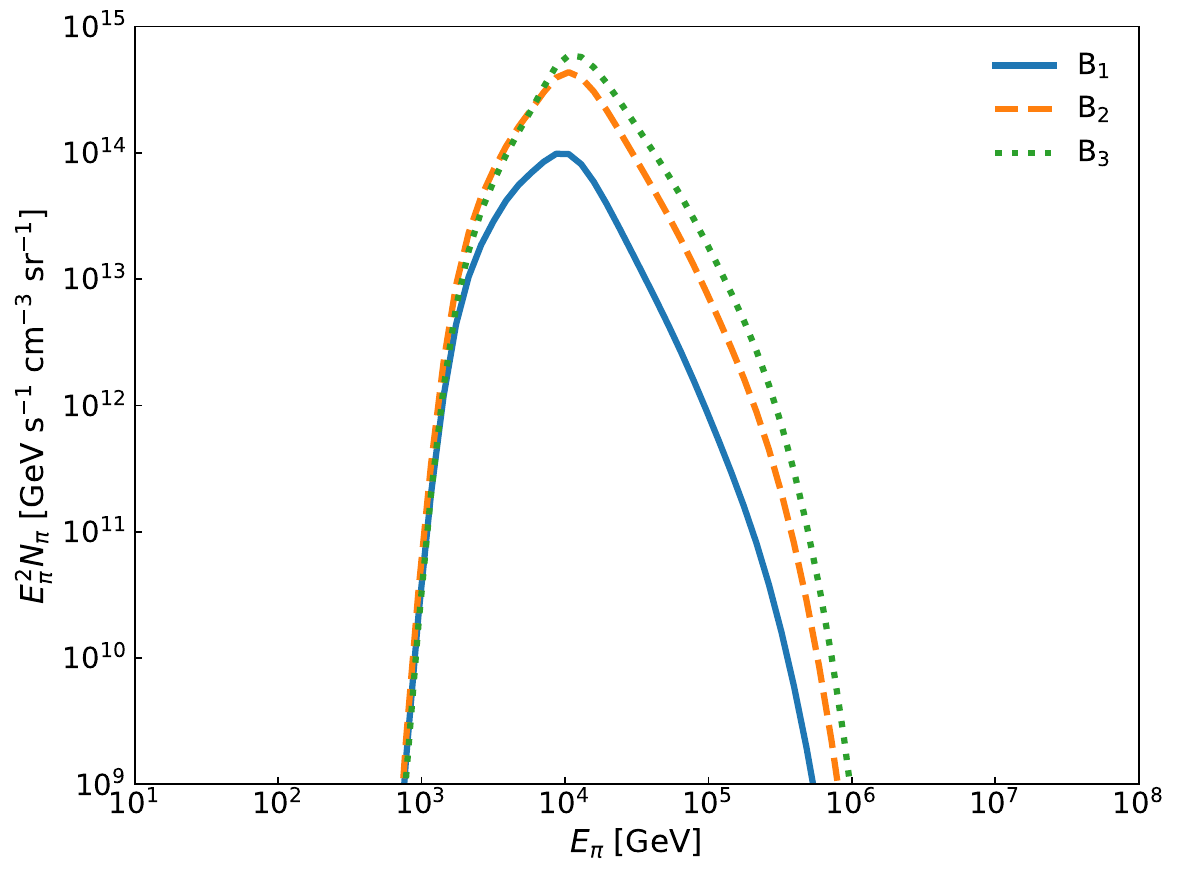} 
    \end{subfigure}
 \caption{Pion distributions for the different parameter sets. The left panel shows the distributions for the $A_i$ scenarios, which reach a maximum at $E_\pi \sim$TeV, while the right panel shows the $B_i$ scenarios, which exhibit a lower peak at $E_\pi \sim 10\,\mathrm{TeV}$}
\label{fig:pi-distributions}
\end{figure*}

\begin{figure*}[t]                            
\centering
\  \centering
    \begin{subfigure}[t]{0.49\textwidth}
        \centering            \includegraphics[width=0.5\linewidth,trim= 115 30 115 20]{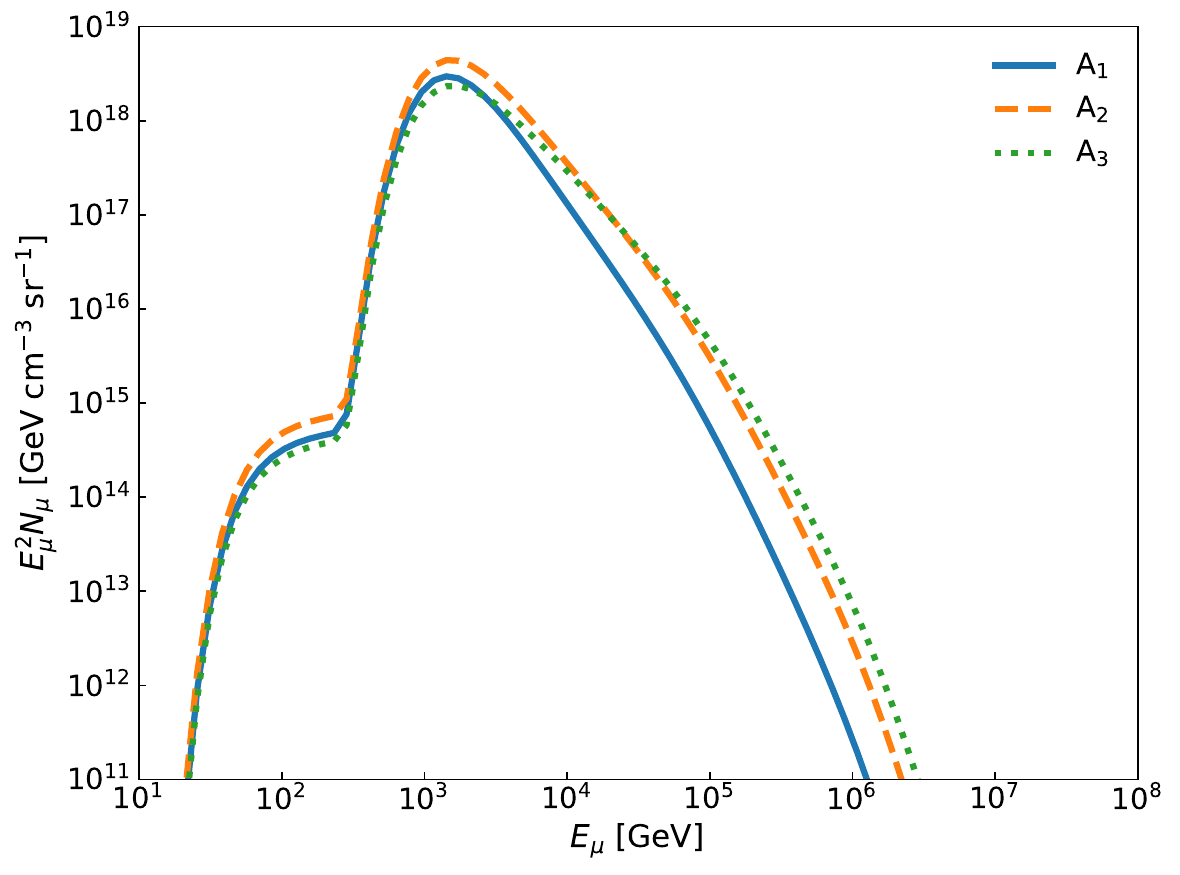} 
    \end{subfigure}
    \hfill
    \begin{subfigure}[t]{0.49\textwidth}
        \centering
    \includegraphics[width=0.5\linewidth,trim= 115 30 115 30]{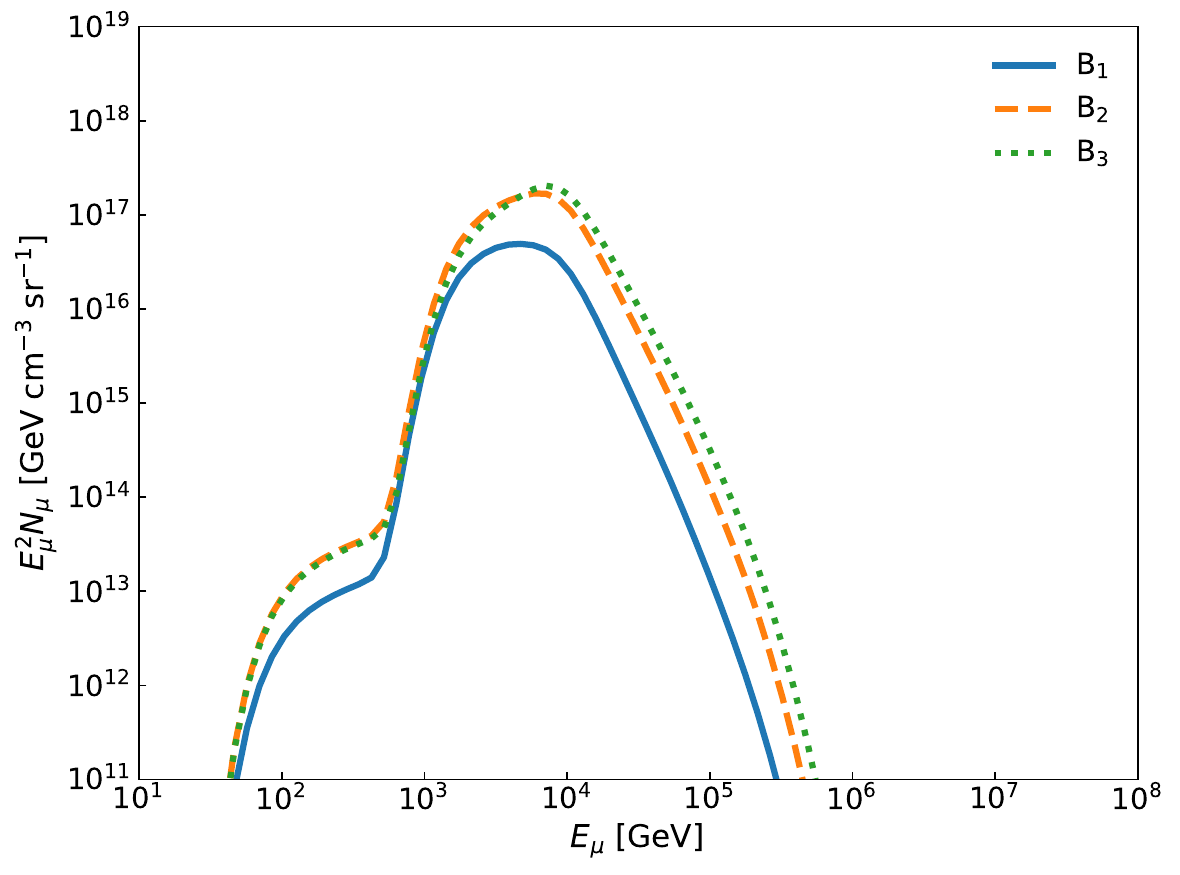} 
    \end{subfigure} 
 \caption{Muon distributions for the different parameter sets. The left panel shows the distributions for the $A_i$ scenarios, where the flux peaks at $E_\mu \approx$TeV. The right panel shows the distributions for the $B_i$ scenarios, where the peak is lower than in the $A_i$ set and occurs at higher energies, $E_\mu \approx 10\,\mathrm{TeV}$.
}\label{fig:mu-distributions} 

\end{figure*}

The low density inside the funnel suggests that the production of neutrinos will be primarily driven by $p\gamma$ interactions with the disk photon field. In the proton rest frame, for photon energies $E_{\rm ph}>2m_ec^2$ and up to $E_{\rm ph}\sim 145$ MeV, the main channel is pair production via Bethe-Heitler mechanism, $p+\gamma \longrightarrow p+ e^- + e^+$.
From $E_{\rm ph}\sim 145$ MeV onward, photomeson production dominates,
$p+\gamma \longrightarrow p+ a\pi^0 + b(\pi^++\pi^-)$
or
$p+\gamma \longrightarrow n + \pi^++ a\pi^0 + b(\pi^++\pi^-),$
where $a$ and $b$ denote multiplicities of neutral and charged pions. Charged pions decay into leptons and neutrinos 
$$\pi^+\longrightarrow e^+ + \nu_\mu + \nu_e +\bar{\nu}_\mu,$$
$$\pi^-\longrightarrow e^- + \bar{\nu}_\mu + \bar{\nu}_e + \nu_\mu,$$
while neutral pions decay into gamma rays
$\pi^0\longrightarrow2\gamma.$


We obtain the particle distributions by solving a stationary one-zone transport equation for each particle species $i=\left\{p,\pi^\pm,\mu^\pm \right\}$ \citep{khangulyan2007}:
\begin{equation}
    \frac{\mathrm{d}\left[b_{i,\rm loss}(E_i) N_i(E_i)\right]}{\mathrm{d}E_i}+  \frac{N_i(E_i)}{T_{i,\rm esc}}= Q_i(E_i),\label{Eq_transport_i}  
\end{equation}
where the total energy loss reads
\begin{equation}
    b_{i,\rm loss}= -\left.\frac{\mathrm{d}E_i}{\mathrm{d}t}\right|_{\rm loss}\equiv  E_i \sum_j t^{-1}_{i,j}(E_{i}),
\end{equation}
including the contribution of all the relevant processes, each with a cooling rate $t^{-1}_{i,j}$. The resulting proton distributions obtained for all sets $A_i$ and $B_i$ are shown in Fig. \ref{fig:p-distributions}. 

The shape of the distributions reflects the dominance of different cooling channels across distinct energy ranges. At low energies the spectrum is left unchanged by $pp$ losses. Around 100 GeV, the Bethe–Heitler process starts to prevail, causing a softening of the spectrum before it hardens again. At $\sim$10 TeV, photopion production becomes the main cooling mechanism, leading to another softening followed by a hardening near 100 TeV. Finally, the distributions exhibit a cutoff at the maximum proton energies.

Likewise, the pion injection term $Q_{\pi}$ is computed for $p\gamma$ interactions using the fitted formulae of \cite{2011PhRvD..83f7303B}, which are derived from the SOPHIA code. For pions and muons, an additional decay term is included on the left member of the transport equation, ${N_i}/{T_{i,\rm dec}}$, 
where $T_{\pi,\rm dec}=2.6\times 10^{-8}\gamma_\pi$, and $T_{\mu,\rm dec}=2.2\times 10^{-6}\gamma_\mu$ are the corresponding decay timescales as a function of the Lorentz factor $\gamma_i=E_i/m_i c^2$.

The cooling rates for pions and muons are shown in Figs. \ref{fig:pi-rates} and \ref{fig:mu-rates}, respectively. For pions, we account for $\pi\gamma$ interactions, following the discussions in \cite{lipari2007} and \cite{reynosodeus2023}. For muons, we compute IC losses of muons ($\mu+\gamma\rightarrow\mu+\gamma$) moving through the ambient photon field produced by the disk; other target fields, such as the synchrotron photon field or the thermal radiation from the wind, provide a negligible contribution. In the $A_i$ scenarios, synchrotron losses dominate above $E_\mu\approx100$ GeV. In contrast, for the $B_i$ parameter sets, IC losses dominate up to $E_\mu \approx 100$ GeV, and synchrotron losses become dominant at higher energies. The resulting energy distributions of charged pions are presented in Fig. \ref{fig:pi-distributions}.

The injection spectra of secondary muons are calculated using the expressions from \cite{lipari2007}. The resulting muon distributions are shown in Fig. \ref{fig:mu-distributions}.

\section{Neutrino fluxes} \label{sec: neutrino fluxes}

Having computed the pion and muon distributions, we can obtain the neutrino plus antineutrino emissivities for the electron and muon flavors, $l=\{e,\mu\}$,
\begin{equation}
Q_{\nu_l+\bar{\nu}_l}(E_\nu)
\equiv
\frac{\mathrm{d}\mathcal{N}_{\nu_l}}{\mathrm{d}E_\nu\,\mathrm{d}V\,\mathrm{d}\Omega\,\mathrm{d}t}.\label{eq.Qnudef}
\end{equation}
In the following, we use the term ``neutrinos” for brevity to denote both neutrinos and antineutrinos, as implied by the definition above. These emissivities are computed following \cite{2008MNRAS.387.1745R}, using the decay kinematics of pions and muons as given in \cite{lipari2007}.
It is worth remarking that in astrophysical hadronic sources, neutrino production at the source is in general dominated by pion and muon decays, yielding only electron and muon neutrinos. Tau neutrino production at the source is expected to be negligible, and their detection at Earth arises predominantly from flavor oscillations during propagation (see e.g. \citealt{learned1995,deyoung2007}). Consequently, most treatments of astrophysical neutrino emission typically focus on the relative production of electron and muon neutrinos at the source, which may be affected by synchrotron cooling in magnetized environments \citep[e.g.][]{kashti2005,lipari2007,Reynoso_Romero_Magneticfield_2009,baerwald2011}.

In addition, the detection of astrophysical neutrinos is most efficient in the muon neutrino channel due to the long range of the produced muons in neutrino telescopes. Therefore, we focus on the muon neutrino flux arriving at Earth from a ULX located at a distance $d_{\rm ULX}$:
\begin{eqnarray}
\phi_{\nu_\mu+\bar{\nu}_\mu}(E_\nu)
&\equiv&
\frac{\mathrm{d}\mathcal{N}_{\nu_\mu+\bar{\nu}_\mu}}{\mathrm{d}E_\nu\,\mathrm{d}A\,\mathrm{d}t} \\
&=&
\phi^{(0)}_{\nu_\mu+\bar{\nu}_\mu}(E_\nu)\,P_{\mu\mu}
+
\phi^{(0)}_{\nu_e+\bar{\nu}_e}(E_\nu)\,P_{e\mu},
\label{eq.phinumu0}
\end{eqnarray}
where $\phi^{(0)}_{\nu_l+\bar{\nu}_l}$ denote the emitted (unoscillated) neutrino fluxes at the source, and $P_{l\,l'}$ are the flavor-transition probabilities over astrophysical distances. These are given by
\begin{equation}
P_{l\,l'}
=
\sum_{i=1}^{3}
|U_{l\, i}|^2\,|U_{l'\, i}|^2,
\end{equation}
where $U$ is the neutrino mixing matrix, evaluated using the best-fit oscillation parameters of \cite{esteban2020}.

The emitted neutrino fluxes $\phi^{(0)}_{\nu_l+\bar{\nu}_l}$ are related to the emissivities by assuming isotropic and homogeneous neutrino production within the compact region of volume $\Delta V$ corresponding to the interaction volume considered above. Under these assumptions, it is found that (see \ref{app:flux})
\begin{equation}
\phi^{(0)}_{\nu_l+\bar{\nu}_l}(E_\nu)
=
\frac{\Delta V}{d_{\rm ULX}^2}
\,Q_{\nu_l+\bar{\nu}_l}(E_\nu),
\end{equation}
and substituting into Eq.~(\ref{eq.phinumu0}) yields the following expression for the muon neutrino flux expected at Earth:
\begin{equation}
\phi_{\nu_\mu+\bar{\nu}_\mu}(E_\nu)
=
\frac{\Delta V}{d_{\rm ULX}^2}
\left(
Q_{\nu_\mu+\bar{\nu}_\mu}\,P_{\mu\mu}
+
Q_{\nu_e+\bar{\nu}_e}\,P_{e\mu}
\right)
\;[{\rm GeV^{-1}\,cm^{-2}\,s^{-1}}].
\end{equation}
%
%
%
%


\begin{figure*}[t]                            
\centering
\  \centering
    \begin{subfigure}[t]{0.49\textwidth}
        \centering            \includegraphics[width=0.5\linewidth,trim= 110 30 110 20]{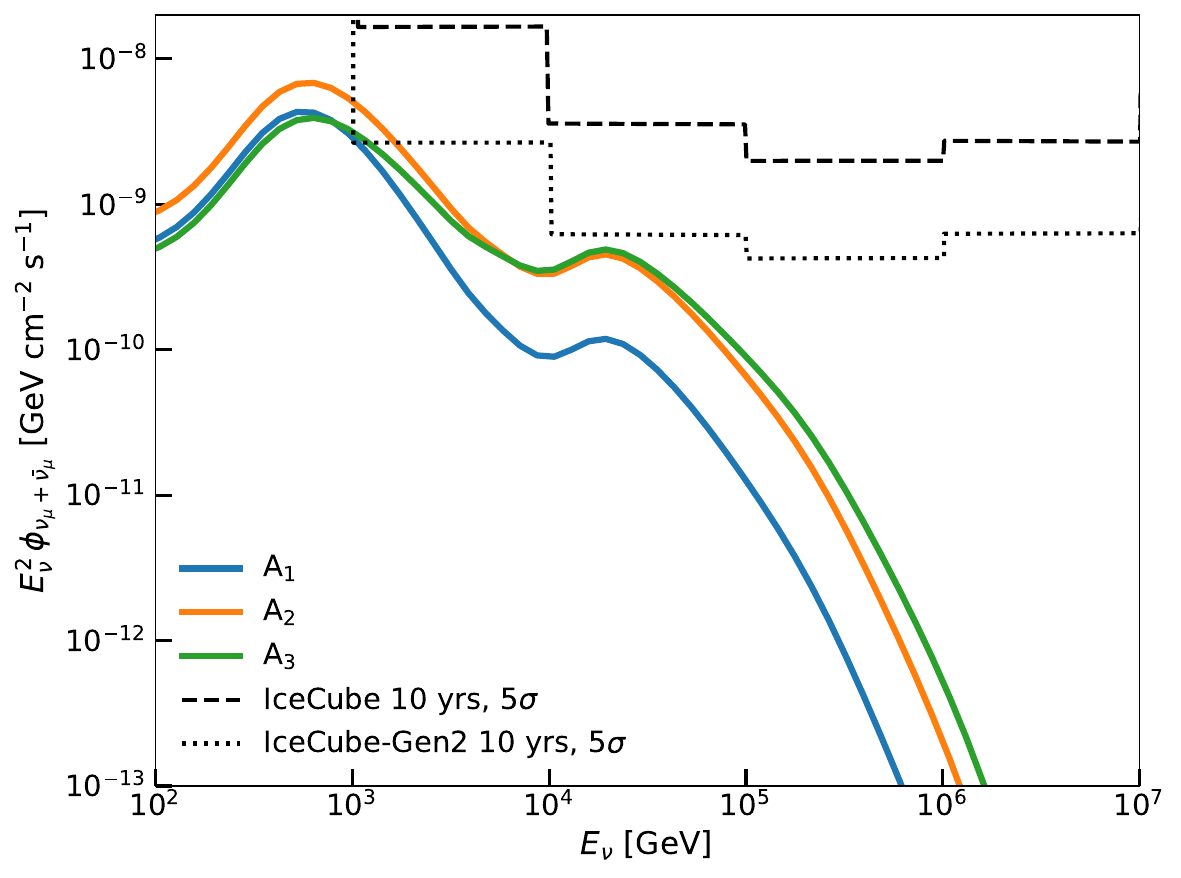} 
    \end{subfigure}
    \hfill
    \begin{subfigure}[t]{0.49\textwidth}
        \centering  \includegraphics[width=0.5\linewidth,trim= 110 30 110 30]{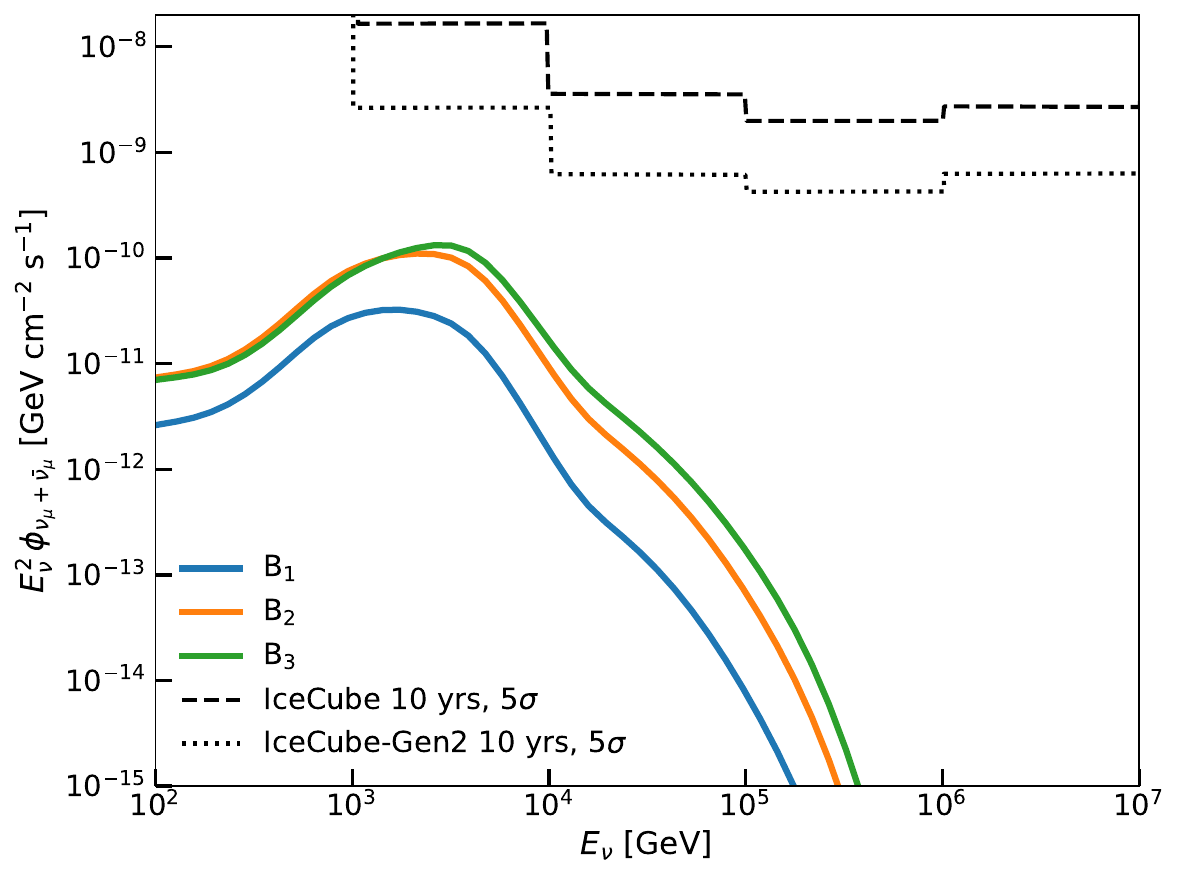} 
    \end{subfigure}
 \caption{Muon neutrinos and antineutrino fluxes for a ULX at $d_{\rm ULX}=10\,{\rm kpc}$. The 10-year sensitivities of IceCube and IceCube-Gen2 are also shown.
 Left: neutrino fluxes for the $A_i$ scenarios, which display a main peak at $E_\nu \sim$ TeV and a secondary peak at $E_\nu \sim 20$ TeV at lower flux levels.
Right: neutrino fluxes for the $B_i$ scenarios, exhibiting a single prominent peak at $E_\nu \approx 2$ TeV, with fluxes roughly one order of magnitude lower than in the $A_i$ sets.}
 \label{fig:numu-fluxes}

\end{figure*}

\begin{figure}[h!]
    \centering   \includegraphics[width=0.95\linewidth,trim= 15 30 15 20]{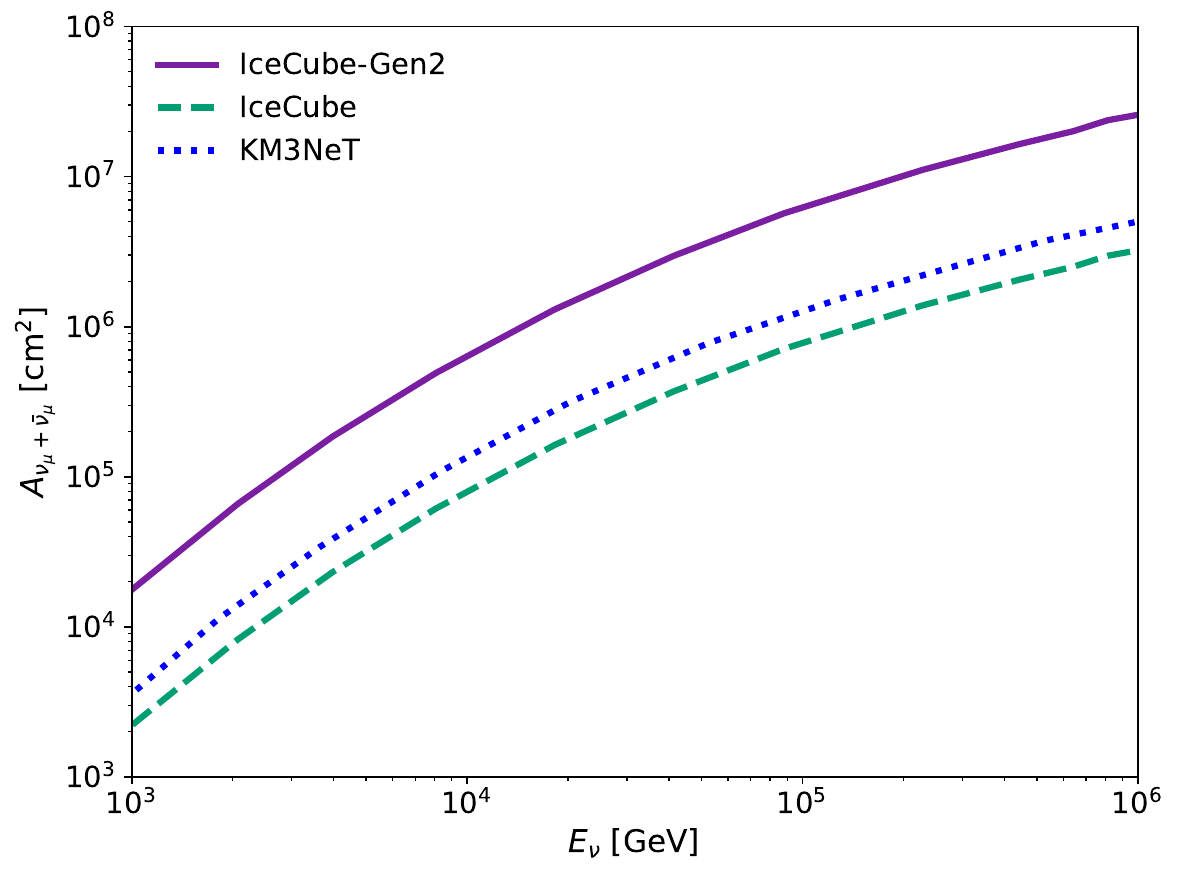}
 \caption{Neutrino effective areas for muon neutrinos plus antineutrinos for IceCube \citep{icecube2025}, IceCube-Gen2 \citep{ladneha_icecubegen2,IceCube_Gen2_2021}, and KM3NeT/ARCA \citep{adrianmartinez2016}.}
 \label{fig.Anueff}
\end{figure}

\begin{figure*}[t]                            
\centering
\  \centering
    \begin{subfigure}[t]{0.49\textwidth}
        \centering            \includegraphics[width=0.5\linewidth,trim= 120 30 120 20]{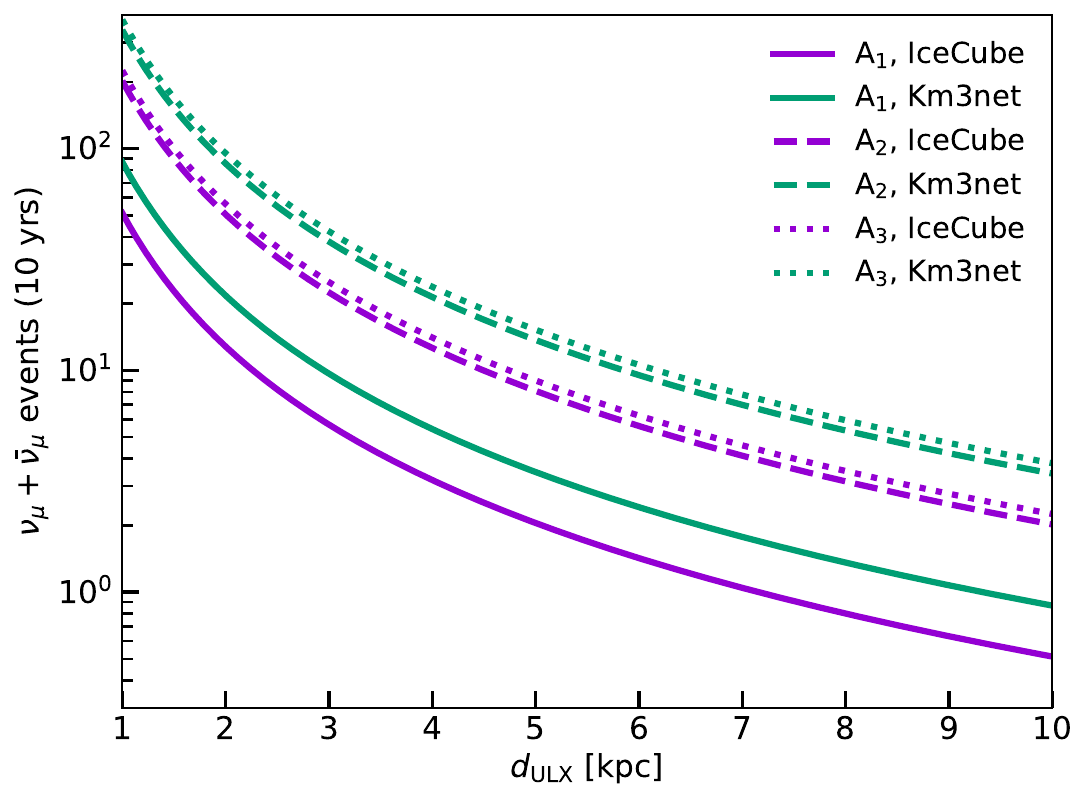} 
    \end{subfigure}
    \hfill
    \begin{subfigure}[t]{0.49\textwidth}
        \centering  \includegraphics[width=0.5\linewidth,trim= 120 30 120 30]{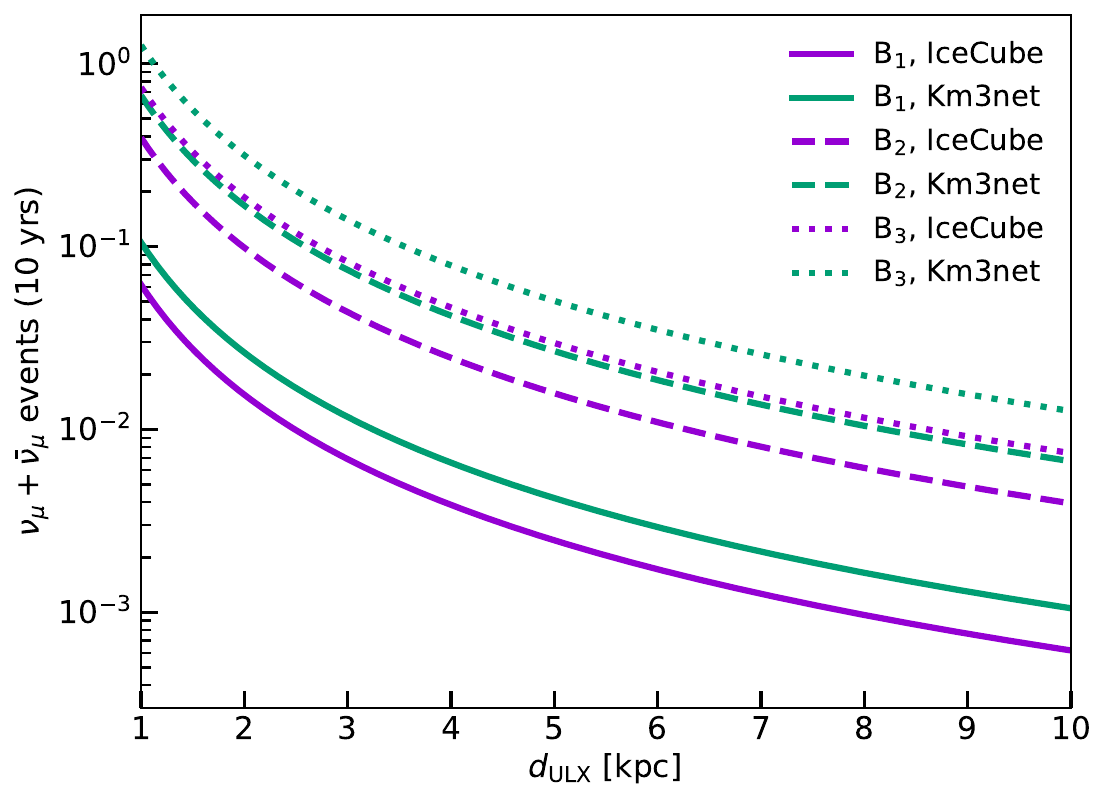} 
    \end{subfigure}
 \caption{Expected number of $\nu_\mu+\bar{\nu}_\mu$ events vs ULX distance to be observed by different detectors in 10 yr of operation.}\label{fig:numu-events}
\end{figure*}

Figure \ref{fig:numu-fluxes} shows the resulting neutrino fluxes for a hypothetical Galactic ULX at a distance of $d_{\rm ULX}=10\,{\rm kpc}$. The only difference between the $i$ scenarios is the spectral index $\Gamma$ of the injected protons, which does not lead to significant qualitative differences among the predicted neutrino fluxes. We note that the thick disk wind formed around the black hole would completely obscure such a system in X-rays when viewed from the side (i.e., $i_{\rm los}\gtrsim 15^\circ$). Even in this obscured scenario, detecting the produced neutrino fluxes remains challenging. 

The number of muon neutrino and antineutrino events per year in the energy range between $E_{\nu,\rm min}=2500\,{\rm GeV}$ and $E_{\nu,\rm max}=10^7{\,\rm GeV}$ is
\begin{equation}
\dot{\mathcal{N}}_{\nu_\mu+\bar{\nu}_\mu}= \int_{E_{\nu,\rm min}}^{E_{\nu,{\rm max}}}\mathrm{d}E_\nu \,A_{\nu, {\rm eff}}(E_\nu)\phi_\nu(E_\nu),
\end{equation}
where $A_{\nu, {\rm eff}}(E_\nu)$ is the effective area of the neutrino detector, which depends on the experimental sampling and event selection techniques. For IceCube, the observation of northern sky sources is favored due to a lower atmospheric background. This corresponds to declinations $\delta > -15^\circ$, where the muon track selection technique is applied. The corresponding effective area \citep{IceCube_Gen2_2021,icecube2025} is shown in Fig.~\ref{fig.Anueff}. The figure also includes the effective areas for the future upgrade IceCube-Gen2 \citep{ladneha_icecubegen2} and for KM3NeT/ARCA \cite{adrianmartinez2016}, which is currently under construction.

The event rates obtained for each parameter set—assuming a source distance of $d=10\,{\rm kpc}$—are presented in Table~\ref{table:events.10kpc}. Additionally, Fig.~\ref{fig:numu-events} shows the expected number of events over a 10-year observation period as a function of the source distance. We see that for some models with parameters from configurations $A_i$ up to 100 events can be expected on timescales of 10 yr for sources at distances $\sim2-3$ kpc. 


\begin{figure}[h!]
    \centering   \includegraphics[width=0.9\linewidth]{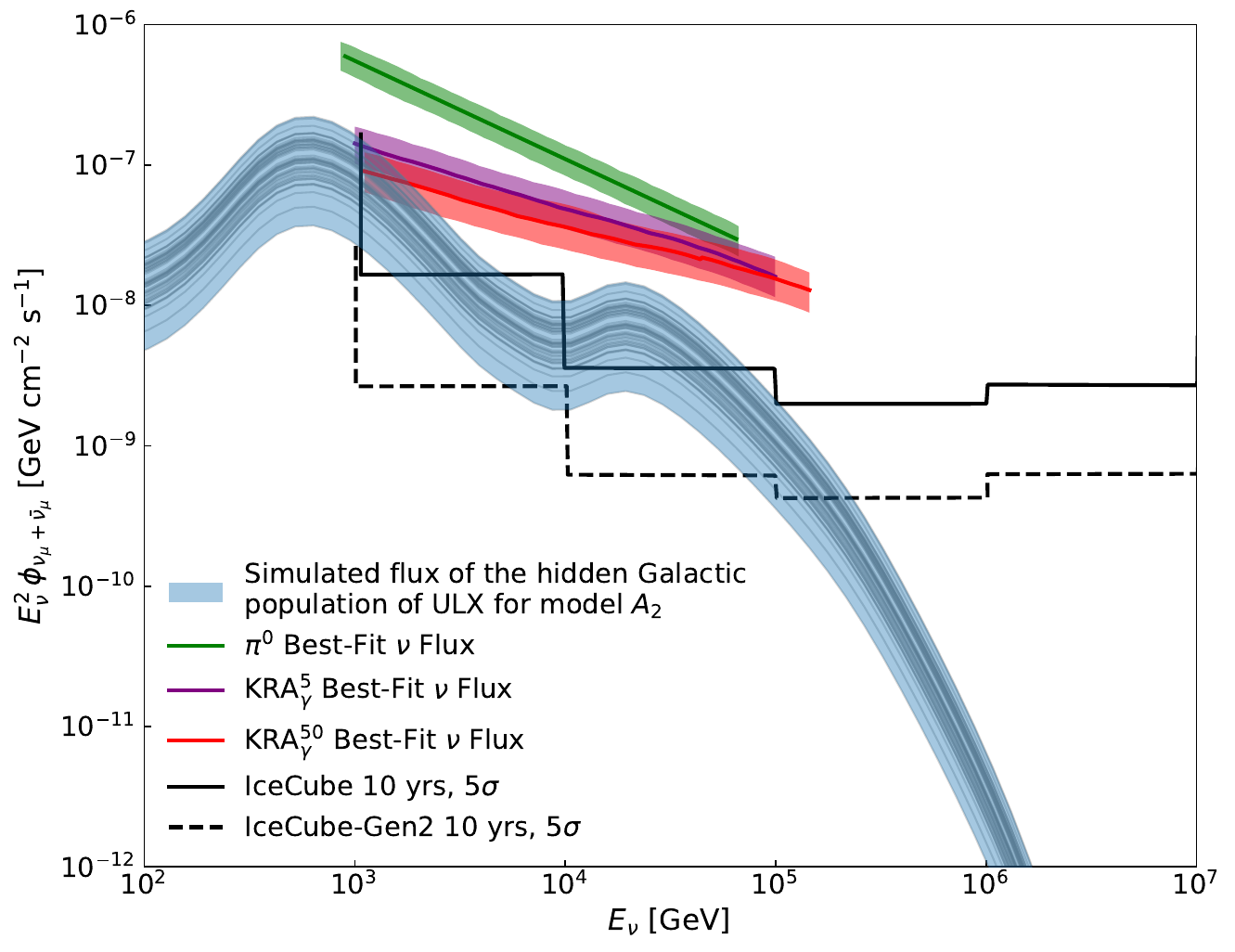}
 \caption{Neutrino spectral energy distributions obtained from  simulated random spatial distributions of seven hidden Galactic ULXs corresponding to the $A_2$ parameter set. The red shaded region encloses the total flux range obtained from all samples. Solid green, purple, and red lines are three possible best-fitting fluxes given by IceCube consistent with the total observed Galactic data \citep{IceCube_galactic_neutrino_2023}. The corresponding  shaded bands indicate the 1$\sigma$ uncertainties. 
 The IceCube discovery potential curves for 10 years of observation are also shown.}
 \label{fig:estimative_flux}
\end{figure}


To assess the possible total contribution of hidden ULXs to the Galactic neutrino flux, we computed the 
individual and collective spectral energy distributions from a
population of seven sources randomly distributed throughout the Galaxy. The results are shown in Figure \ref{fig:estimative_flux} for the optimistic parameter set $A_2$. The red shaded region is the range of total fluxes from all simulated populations. If hidden ULXs are similar to those characterized by model $A_2$, their actual contribution to the neutrino flux should lie within this range. For comparison, we also include three models of the total Galactic neutrino flux observed by IceCube, all of which are consistent with gamma-ray data above $100\,{\rm GeV}$ but differ in their assumptions regarding the parent proton spectra and the spatial distribution of sources in the Galaxy \citep{IceCube_galactic_neutrino_2023}. Our results suggest that the cumulative neutrino emission from Galactic ULXs could account for a significant fraction of the Galactic flux at energies below $\sim 50\,{\rm TeV}$.



\begin{table}[h!]
{\small 
\caption{$\nu_{\mu}+\bar{\nu}_\mu$ events per yr, $\dot{\mathcal{N}}_{\nu_\mu+\bar{\nu}_\mu}[{\rm yr}^{-1}]$, for $d_{\rm ULX}=10\,{\rm kpc}$}\label{table:events.10kpc}
\centering                                      
\begin{tabular}{l c c c}          
\hline                     
parameter set &  IceCube  & IceCube-Gen2 &  KM3NeT\\    
\hline                                   
    $A_1$ & $5.1\times 10^{-2}$  & $4.0\times 10^{-1}$ & $8.6\times 10^{-2}$ \\
    $A_2$ & $2.0\times 10^{-1}$  & $1.6$ & $3.4\times 10^{-1}$ \\  
 $A_3$ & $2.2\times 10^{-1}$  & $1.8$ & $3.8\times 10^{-1}$ \\
$B_1$ & $6.2\times 10^{-5}$  & $4.9\times 10^{-4}$ & $1.0\times 10^{-4}$ \\
$B_2$ & $3.9\times 10^{-4}$  & $3.5\times 10^{-3}$ & $6.6\times 10^{-4}$
\\
$B_3$ & $7.4\times 10^{-4}$  & $5.9\times 10^{-3}$ & $1.2\times 10^{-3}$  \\ 
\hline
\end{tabular}
}
\end{table}

\section{Discussion}\label{sec: discussion}


The primary difference between the $A_i$ and $B_i$ scenarios lies in the accretion rate, which significantly impacts the structure of both the accretion disk and the wind. The disk's physical properties are determined by key parameters—such as the viscosity $\alpha$, the adiabatic index $\gamma$, the advection parameter $f$, and the magnetization $\beta$ \citep{akizuki2006}—for which we assume values typical of a supercritical disk. The resulting magnetic field and disk temperature, in turn, affect the physical conditions in the acceleration region. In contrast, the wind temperature is not a key parameter, as the radiation field from the disk dominates within the funnel.

The accretion rate is also related to the semi-opening angle of the funnel formed by the wind walls. Although the exact functional dependence is unknown, higher accretion rates are expected to produce stronger outflows that regulate accretion onto the black hole at the Eddington limit, thereby reducing the funnel size. For this reason, we explore scenarios with semi-opening angles of $15^\circ$ ($A_i$), and $10^\circ$ ($B_i$), which directly affect the power available to accelerate particles. In the former case, the available power is approximately three orders of magnitude higher than in the latter, which has a direct impact on the predicted neutrino fluxes and the expected number of observable events.

Our model does not address the specific mechanism of particle injection into the funnel. Simulations of supercritical accretion flows suggest that non-dipolar magnetic fields can naturally channel part of the disk material toward the axis, favoring episodic mass loading \citep{McKinney2009,Romero2021AN}. Conversely, modeling magnetic reconnection requires simplifying assumptions. The small-scale geometry of magnetic fields in supercritical disks plays a key role in controlling mass loading and determining the funnel's internal structure. Future numerical simulations, combined with high-resolution observations, will be essential to determine whether such injection channels and reconnection processes do occur in nature. Rapid variability in the hard X-rays may be indicative of reconnecting activity at the base of the funnel.   

The model proposed by \citet{NSULX_neutrinos_2025} for the subclass of ULXs known as PULXs (pulsating ULXs) indicates that sources with $L_{X} > 10^{39}$ erg s$^{-1}$ can act as efficient neutrino emitters. These authors compared their predictions with the Galactic PULX Swift J0243.6+6124, located at a distance of 5--7 kpc \citep{2018ApJ...863....9W, 2020A&A...640A..35R}. Although no neutrinos from this source have been detected by IceCube, their results suggest that identifying neutrino signals from transient PULXs with $L_{X}\!\approx\!5\times10^{39}$ erg s$^{-1}$, such as Swift J0243.6+6124, becomes increasingly difficult beyond $\sim$4 kpc. Nonetheless, neutron-star ULXs might also produce TeV neutrinos through magnetospheric acceleration \citep[e.g.][]{Anchordoqui2003ApJ}.

An important implication of our results is that misaligned ULXs may remain undetectable in X-rays while still contributing to the Galactic neutrino background. The cumulative emission from such hidden systems could account for a non-negligible fraction of the diffuse Galactic neutrino flux reported by IceCube \citep{2024ApJ...969..161G}. Although only a handful of super-Eddington X-ray binaries are currently known in the Milky Way, population-synthesis studies combined with geometrical obscuration effects suggest that the total number could be on the order of $5$--$10$, with about a half hosting a black hole in low solar metallicity environments \citep[e.g.][]{Wiktorowicz2019ApJ}. Observational surveys of ULXs in nearby galaxies support this estimate: \citet{Kovlakas2020} find an occurrence rate of about $0.45 \, {\rm ULX}\, (M_\odot\,{\rm yr}^{-1})^{-1}$. For a Galactic SFR of $1$--$3\,M_\odot\,{\rm yr}^{-1}$, this implies $\sim 0.5$--$1.5$ observable ULXs, which rises to several when accounting for sources hidden by unfavorable beaming or absorption (see also \citealt{Grimm2003}). Under optimistic assumptions, the total Galactic population could therefore reach $\sim 5$--$10$. This hidden population may thus constitute a significant collective contributor to the Galactic neutrino flux. 

Finally, we note that significant internal $\gamma$-ray absorption is expected due to $\gamma\gamma$ annihilation with the disk's dense photon fields or, in the case of misaligned, obscured ULXs, due to $\gamma n$ absorption within the dense wind \citep{Romero-Pasqa-Abaroa2025}. Consequently, such sources would be electromagnetically quiet at high energies yet still produce detectable neutrino fluxes.






\section{Conclusions} \label{sec: conclusion}

We have investigated neutrino production in Galactic ULXs powered by stellar-mass black holes accreting at super-Eddington rates. Our results indicate that these sources can be efficient sites of hadronic acceleration and neutrino emission. In the explored parameter space, protons reach energies in the PeV range for moderate accretion regimes ($\dot{m}=10$), whereas in hyperaccreting cases ($\dot{m}=10^3$), the maximum energy is limited to $\sim 100$ TeV due to severe radiative cooling. In both scenarios, $p\gamma$ interactions with the X-ray photon field of the inner disk are the dominant interactions and produce a neutrino flux by pion and muon decay. We modeled six parameter sets for both accretion regimes, corresponding to different injection indices for hadrons ($\Gamma=2,1.5,1$) and progressively lower mass densities.

 Misaligned ULXs, obscured in X-rays by optically thick winds, can remain hidden in the X-ray band while still contributing to the diffuse Galactic neutrino background. These sources might represent a missing component in multimessenger studies of the Galaxy.

The predicted muon neutrino fluxes shown in Fig. \ref{fig:numu-fluxes} for a source located at $d_{\rm ULX}=10$ kpc suggest that the planned IceCube-Gen2 observatory will have the capability to detect neutrinos from standard ULXs within the Milky Way. More nearby, obscure sources, are detectable even by IceCube with its current configuration. 

Future searches combining high-energy neutrinos with optical, X-ray, and gamma-ray surveys will be crucial to identifying these systems. A detection would not only establish ULXs as hadronic accelerators but also provide new insights into the physics of supercritical accretion and the demographics of hidden compact objects in our Galaxy.

\section*{Acknowledgements}
 GER was funded by PID2022-136828NB-C41/AEI/10.13039/501100011033/ and through the ``Unit of Excellence María de Maeztu'' award to the Institute of Cosmos Sciences (CEX2019-000918-M, CEX2024-001451-M). Additional support came from PIP 0554 (CONICET). MMR acknowledges support from UNMdP through grant 80020240500217MP.

\appendix

\appendix
\section{Relation between neutrino emissivities and emitted flux}
\label{app:flux}

In this Appendix we derive the relation between the neutrino emissivities at
the source and the corresponding emitted (unoscillated) neutrino fluxes. The starting point is the neutrino emissivity of flavor
$l=\{e,\mu\}$ as defined in Eq. (\ref{eq.Qnudef}),
which represents the number of neutrinos produced per unit energy, volume,
solid angle, and time. And we aim to link this to the emitted neutrino flux defined as
\begin{equation}
\phi^{(0)}_{\nu_l+\bar{\nu}_l}(E_\nu)
\equiv
\left.
\frac{\mathrm{d}\mathcal{N}_{\nu_l+\bar{\nu}_l}}{\mathrm{d}E_\nu\, \mathrm{d}A\, \mathrm{d}t}
\right|_{\rm emitted},
\end{equation}
i.e. the number of neutrinos crossing a detector area element $\mathrm{d}A$ per unit
energy and time, before flavor oscillations are taken into account.

The two quantities are related through the geometrical relation between the
solid-angle and area elements. For isotropic emission from a source located
at a distance $d_{\rm ULX}$,
\begin{equation}
\frac{\mathrm{d}\Omega}{\mathrm{d}A} = \frac{1}{d_{\rm ULX}^2},
\end{equation}
which reflects the fact that neutrinos arriving at a differential area
element $dA$ must have been emitted within a solid-angle element
$d\Omega = dA/d_{\rm ULX}^2$. Using this relation, the emitted flux can be written as
\begin{equation}
\phi^{(0)}_{\nu_l+\bar{\nu}_l}= \frac{\mathrm{d}\mathcal{N}_{\nu_l+\bar{\nu}_l}}{\mathrm{d}E_\nu\, \mathrm{d}\Omega\, \mathrm{d}t}
\left(\frac{\mathrm{d}\Omega}{\mathrm{d}A}\right).
\end{equation}

The connection with the emissivities is then obtained by integrating over the
neutrino production volume $\Delta V$,
\begin{equation}
\phi^{(0)}_{\nu_l+\bar{\nu}_l}
=
\int_{\Delta V} dV\,
Q_{\nu_l+\bar{\nu}_l}(E_\nu)\,
\frac{1}{d_{\rm ULX}^2}.
\end{equation}

Assuming homogeneous and isotropic neutrino production within the volume
$\Delta V$, the emissivities are constant inside this region and the
expression reduces to
\begin{equation}
\phi^{(0)}_{\nu_l+\bar{\nu}_l}(E_\nu)
=
\frac{\Delta V}{d_{\rm ULX}^2}
\, Q_{\nu_l+\bar{\nu}_l}(E_\nu),
\end{equation}
which is the relation used in the main text.


\bibliographystyle{elsarticle-harv} 
\bibliography{example}






\end{document}